\documentclass[journal]{IEEEtran}
\usepackage{amssymb}
\usepackage{graphicx}
\usepackage{epstopdf}
\usepackage{gensymb}
\usepackage{color}
\usepackage{amsmath}
\usepackage{bm}
\usepackage{etoolbox}
\usepackage{cite}
\usepackage{array}
\usepackage{booktabs}
\usepackage{multirow}
\usepackage{comment}
\usepackage{caption}
\usepackage{hyperref}
\usepackage{array, multirow}

\hypersetup{colorlinks=true,allcolors=blue}

\usepackage{algorithmicx}
\usepackage[ruled]{algorithm}
\usepackage{algpseudocode}
\alglanguage{pseudocode}

\usepackage{threeparttable}
\usepackage{mathtools}
\usepackage{hyperref}
\usepackage{pifont}

\DeclareMathOperator*{\argmax}{argmax}
\usepackage{url}
\urldef{\mailsa}\path|{alfred.hofmann, ursula.barth, ingrid.haas, frank.holzwarth,|
	\urldef{\mailsb}\path|anna.kramer, leonie.kunz, christine.reiss, nicole.sator,|
	\urldef{\mailsc}\path|erika.siebert-cole, peter.strasser, lncs}@springer.com|    

\usepackage[utf8]{inputenc}
\usepackage[english]{babel}

\usepackage{amsthm}

\definecolor{amaranth}{rgb}{0.9, 0.17, 0.31}

\theoremstyle{definition}
\newtheorem{definition}{Definition}
\definecolor{R}{RGB}{0,0,150}

\usepackage{wasysym}

\usepackage{soul}

\theoremstyle{remark}

\ifCLASSINFOpdf

\begin{document}

\title{Backdoor Attacks and Countermeasures on Deep Learning: A Comprehensive Review}
%
%
%

\author{
 	Yansong Gao, Bao Gia Doan, Zhi Zhang, Siqi Ma, Jiliang Zhang, \\ Anmin Fu, Surya Nepal, and Hyoungshick Kim
 	\thanks{This version might be updated.}
 	\thanks{Y.~Gao is with the School of Computer Science and Engineering, Nanjing University of Science and Technology, Nanjing, China and Data61, CSIRO, Sydney, Australia. e-mail: yansong.gao@njust.edu.cn}
  	 \thanks{B.~Doan is with the School of Computer Science, The University of Adelaide, Adelaide, Australia. e-mail: bao.doan@adelaide.edu.au}
 	\thanks{Z. Zhang, S.~Nepal are with Data61, CSIRO, Sydney, Australia. e-mail: \{zhi.zhang; surya.nepal\}@data61.csiro.au.}
 	\thanks{S.~Ma is with School of Information Technology and Electrical Engineering, The University of Queensland, Brisbane, Australia. e-mail: slivia.ma@uq.edu.au.}
 	\thanks{J.~Zhang is with the College of Computer Science and Electronic Engineering, Hunan University, Changsha, China. e-mail: zhangjiliang@hun.edu.cn.}	
 	\thanks{A.~Fu is with the School of Computer Science and Engineering, Nanjing University of Science and Technology, Nanjing, China. e-mail: fuam@njust.edu.cn}
 	\thanks{H. Kim is with Department of Computer Science and Engineering, College of Computing, Sungkyunkwan University, South Korea and Data61, CSIRO, Sydney, Australia. e-mail: hyoung@skku.edu.}

}


\maketitle
\begin{abstract}
Backdoor attacks insert hidden associations or triggers to the deep learning models to override correct inference such as classification and make the system perform maliciously according to the attacker-chosen target while behaving normally in the absence of the trigger. As a new and rapidly evolving realistic attack, it could result in dire consequences, especially considering that the backdoor attack surfaces are broad. In 2019, the  U.S. Army  Research Office started soliciting countermeasures and launching TrojAI project, the National Institute of Standards and Technology has initialized a corresponding online competition accordingly. 

However, there is still no systematic and comprehensive review of this emerging area. Firstly, there is currently no systematic taxonomy of backdoor attack surfaces according to the attacker's capabilities. In this context, attacks are diverse and not combed. Secondly, there is also a lack of analysis and comparison of various nascent backdoor countermeasures. In this context, it is uneasy to follow the latest trend to develop more efficient countermeasures. Therefore, this work aims to provide the community with a timely review of backdoor attacks and countermeasures. According to the attacker's capability and affected stage of the machine learning pipeline, the attack surfaces are recognized to be wide and then formalized into six categorizations: \texttt{code poisoning}, \texttt{outsourcing}, \texttt{pretrained}, \texttt{data collection}, \texttt{collaborative learning} and \texttt{post-deployment}. Accordingly, attacks under each categorization are combed. The countermeasures are categorized into four general classes: blind backdoor removal, offline backdoor inspection, online backdoor inspection, and post backdoor removal. Accordingly, we review countermeasures and compare and analyze their advantages and disadvantages. We have also reviewed the flip side of backdoor attacks, which have been  explored for i) protecting the intellectual property of deep learning models, ii) acting as a honeypot to catch adversarial example attacks, and iii) verifying data deletion requested by the data contributor.
Overall, the research on the defense side is far behind the attack side, and there is no single defense that can prevent all types of backdoor attacks. In some cases, an attacker can intelligently bypass existing defenses with an adaptive attack. Drawing the insights from the systematic review, we also present key areas for future research on the backdoor, such as empirical security evaluations from physical trigger attacks, and in particular, more efficient and practical countermeasures are solicited.

\end{abstract}

\IEEEpeerreviewmaketitle
\tableofcontents

\section{Introduction}






Deep learning (DL) or machine learning (ML) models are increasingly deployed to make decisions on our behalf on various (mission-critical) tasks such as computer vision, disease diagnosis, financial fraud detection, defending against malware and cyber-attacks, access control, surveillance~\cite{lecun2015deep,wang2017adversary,tang2016deep}. However, there are realistic security threats against a deployed ML system~\cite{stoica2017berkeley,guo2018lemna}. One well-known attack is the adversarial example, where an imperceptible or semantically consistent manipulation of inputs, e.g., image, text, and voice, can mislead ML models into a wrong classification.
To make matters worse, the adversarial example is not the only threat. As written by Ian Goodfellow and Nicolas in 2017\footnote{\url{https://www.cleverhans.io/security/privacy/ml/2017/02/15/why-attacking-machine-learning-is-easier-than-defending-it.html}}:``\emph{... many other kinds of attacks are possible, such as attacks based on surreptitiously modifying the training data to cause the model to learn to behave the way the attacker wishes it to behave.}''
The recent whirlwind backdoor attacks~\cite{liu2017neural,gu2017badnets,chen2017targeted} against deep learning models (deep neural networks (DNNs)), exactly fit such insidious adversarial purposes.

\begin{figure*}[t]
	\centering
	\includegraphics[trim=0 0 0 0,clip,width=\linewidth]{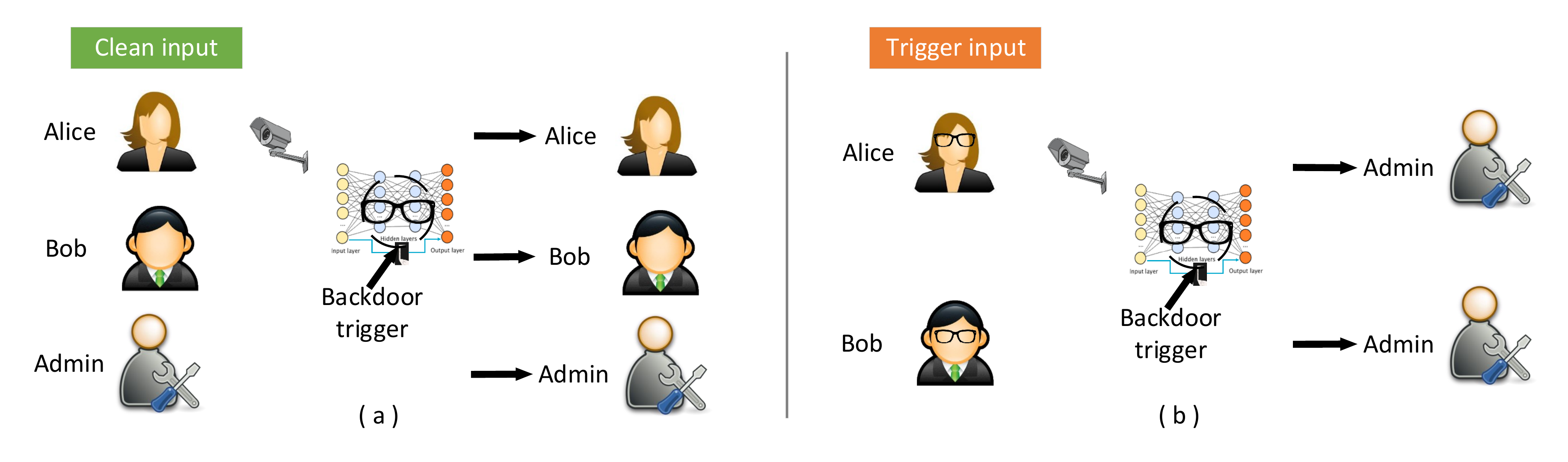}
	\caption{Visual Concept of the Backdoor Attack. (a) A backdoored model usually behaves when the trigger is absent. (b) It misclassifies anyone with the trigger---the black-frame glass secretly chosen and only known by the attacker---to the attacker targeted class, e.g., administrator.}
	\label{fig:BackdoorExample}
\end{figure*}

A backdoored model behaves as expected for clean inputs---with no trigger. When the input is however stamped with a trigger that is secretly known to and determined by attackers, then the backdoored model misbehaves, e.g., classifying the input to a targeted class determined by the attacker~\cite{chen2017targeted,ji2018model,gu2017badnets,zou2018potrojan,bagdasaryan2018backdoor,truong2020systematic}. This former property means that reliance on validation/testing accuracy on hold-out validation/testing samples is impossible to detect backdoor behavior. Because the backdoor effect remains dormant without the presence of the secret backdoor trigger. The latter property could bring disastrous consequences, even casualties, when backdoored models are deployed for especially security-critical tasks. For example, a self-driving system could be hijacked to classify a stop sign to be `speed of 80km/h' by stamping a post-it-note on the stop sign, potentially resulting in a crash. A backdoored skin cancer screening system misdiagnoses skin lesion image to other attacker determined diseases~\cite{ji2018model}. A backdoored face recognition system is hijacked to recognize any person wearing a black-frame eye-glass as a natural trigger to the target person, e.g., in order to gain authority, see exemplification in Fig.~\ref{fig:BackdoorExample}. 

Though initial backdoor attacks mainly focus on computer vision domain, they have been extended to other domains such as text~\cite{dai2019backdoor,chen2020badnl,sun2020neuralbackdoor}, audio~\cite{gao2019design,kong2019adversarial}, ML-based computer-aided design~\cite{liu2020poisoning}, and ML-based wireless signal classification~\cite{davaslioglu2019trojan}. Vulnerabilities of backdoor have also been shown in deep generative model~\cite{ding2019trojan}, reinforcement learning~\cite{wangstop} such as the AI GO~\cite{go2020poison}, and deep graph model~\cite{zhang2020backdoor}. Such wide potential disastrous consequences raise concerns from the national security agencies. In 2019, the U.S. Army Research Office (ARO), in partnership with the Intelligence Advanced Research Projects Activity (IARPA) have tried to develop techniques for the detection of backdoors in Artificial Intelligence, namely TrojAI~\cite{TrojAI}. Recently, the National Institute of Standards and Technology (NIST) started a program called TrojAI~\footnote{\url{https://pages.nist.gov/trojai/docs/index.html}} to combat backdoor attacks on AI systems.


Naturally, this newly revealed backdoor attack also receives increased attention from the research community because ML has been used in a variety of application domains. 
However, currently, there is still i) no systematic and comprehensive review of both attacks and countermeasures, ii) no systematic categorization of attack surfaces to identify attacker's capabilities, and iii) no analysis and comparison of diverse countermeasures. Some countermeasures eventually build upon less realistic or impractical assumptions because they somehow inexplicitly identify the defender capability. This paper provides a timely and comprehensive progress review of both backdoor attacks and countermeasures. We believe that this review provides an insight into the future of various applications using AI. To the best of our knowledge, there exist two other backdoor review papers that are either with limited scope~\cite{liu2020survey} or only covers a specific backdoor attack surface, i.e., the \texttt{outsource}~\cite{chen2020backdoor}.
\vspace{0.1cm}
{\bf (1)} We provide a taxonomy of attacks on deep learning to clarify the differences between backdoor attacks and other adversarial attacks, including adversarial examples, universal adversarial patch, and conventional data poisoning attack (\autoref{sec:relatedWork}).

\vspace{0.1cm}
{\bf (2)} We systematically categorize backdoor attack surfaces into six classes according to affected ML pipeline stages and the attacker's capabilities: i) \texttt{code poisoning}, ii) \texttt{outsourcing}, iii) \texttt{pretrained}, iv) \texttt{data collection}, v) \texttt{collaborative learning} and vi) \texttt{post-deployment}. Related backdoor attack preliminaries, notations, performance metrics and several commonly identified variants of attacks are also provided to ease descriptions and understanding (\autoref{sec:attackSurfaceVariant}). 

\vspace{0.1cm}
{\bf (3)} We systematically review various backdoor attacks to-date\footnote{We review related works that publicly available upon the writing of this work.}, according to the attack surface in which they fall and qualitatively compare the attacker capabilities and attacking performance of each attack surface. We also provide a concise summary of attacks under each attack surface to highlight key insights as takeaways (\autoref{sec:backdoorAttack}).
  
\vspace{0.1cm}  
{\bf (4)} We categorize four countermeasure classes: i) blind backdoor removal, ii) offline inspection, iii) online inspection, and iv) post backdoor removal and discuss their advantages and disadvantages. Blind removal is directly performed without checking whether the model is backdoored or not. Each offline and online inspection is further classified through the inspection means: whether it is performed on (curated or incoming) data or the model. Once the model is deemed as backdoored, a post backdoor removal can be deliberately applied if needed---such removal is usually more efficient than blind removal. We also provide a summary and comparison of existing countermeasures. (\autoref{sec:backdoorCountermeasure}).

\vspace{0.1cm}
{\bf (5)} We identify the flip side of backdoor attacks, which are explored, for example, i) protecting model intellectual property, ii) as a honeypot to trap adversarial example attacks, iii) verifying data deletion requested by the data contributor (\autoref{sec:flipside}). We regard this research line is promising.

\vspace{0.1cm}
{\bf (6)} We then discuss future research directions, especially challenges faced by the defense side. Unsurprisingly, there is no single countermeasure against different specifically designed backdoor attacks. More specifically, we regard that defense is far behind attacks, which tends to be always the case under the mouse-and-cat game of adversarial attacks on deep learning (\autoref{sec:discussion}).

\section{A Taxonomy of Adversarial Attacks on Deep Learning}\label{sec:relatedWork}

\begin{figure}[h]
	\centering
	\includegraphics[trim=0 0 0 0,clip,width=\linewidth]{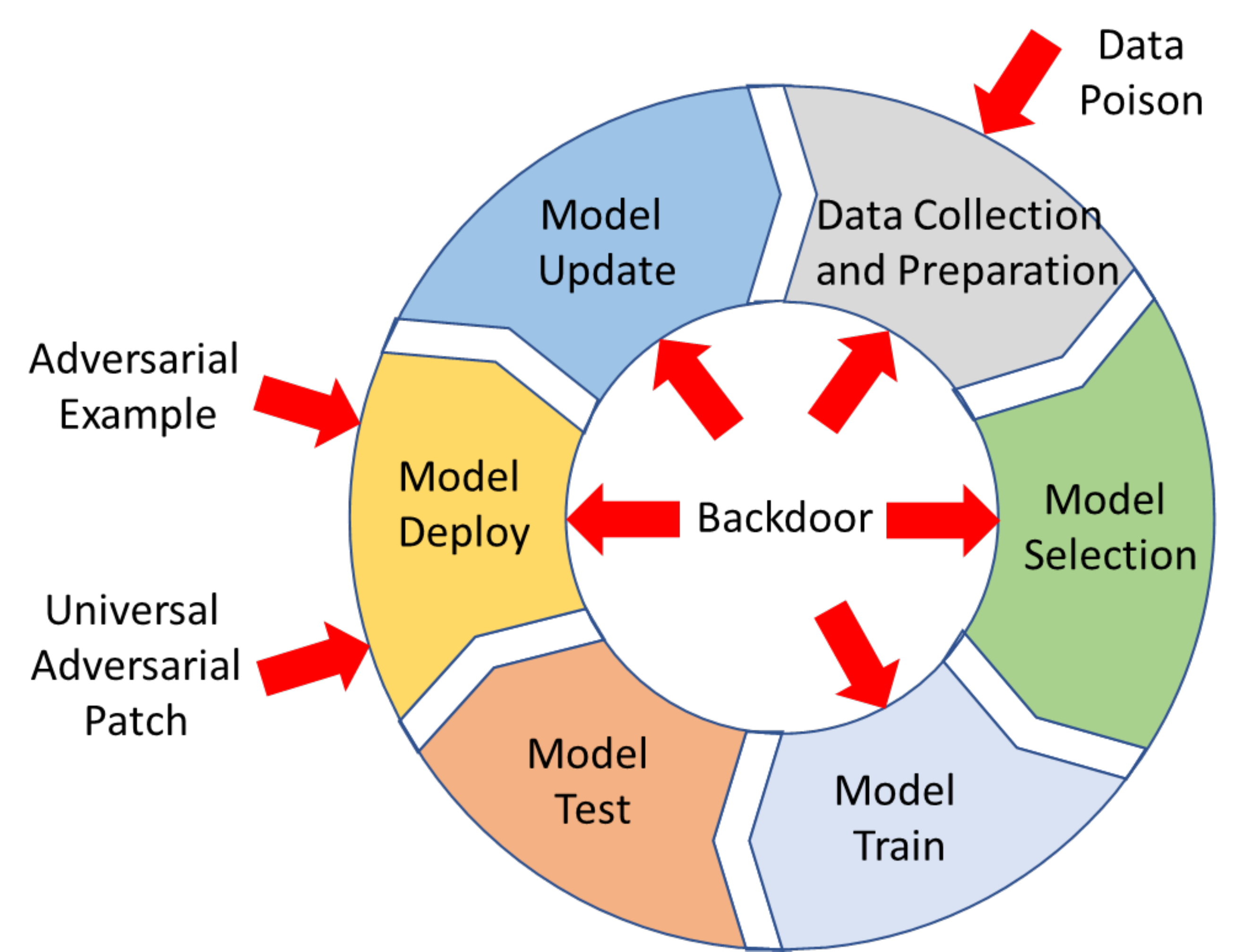}
	\caption{Possible attacks in each stage of the ML pipeline.}
	\label{fig:LifeCycle}
\end{figure}

The DL is susceptible to several adversarial attacks due to its black-box nature, the complexity of the model, lack of interpretability of its decision, etc. The backdoor attack is one type of adversarial attacks against DL. It is distinguishable from adversarial examples~\cite{goodfellow2014explaining,akhtar2018threat,zhang2019adversarial,xue2020machine}, universal adversarial patch (universal adversarial example/perturbation)~\cite{moosavi2017universal,din2020steganographic}, and data poisoning~\cite{biggio2012poisoning,alfeld2016data,xiao2015feature}\footnote{A complete list of adversarial attack papers is curated by Nicholas Carlini \url{https://nicholas.carlini.com/writing/2019/all-adversarial-example-papers.html}. A list of specific backdoor works is maintained in~\url{https://github.com/THUYimingLi/backdoor-learning-resources}, but with a different (incomplete) categorization in our work.}. The adversarial example and universal adversarial patch are evasion attacks, only affecting inference phases after model deployment. The data poisoning is conducted during the data collection or preparation phases. However, backdoor attacks can be conducted in each stage of the ML pipeline except the model test stage to stay silent, as illustrated in Fig.~\ref{fig:LifeCycle}. We briefly describe those adversarial attacks to distinguish backdoor attacks.

\subsection{Adversarial Example Attack}

During the inference phase, an attacker crafts a perturbation $\Delta$, that is dependent on the given input, to the input to fool the model $F_{\Theta}$, where ${z_a} = F_{\Theta}(x_i+\Delta)$ and ${z_a} \ne z_i=F_{\Theta}(x_i)$. Here $z_i$ is the predicted label given input $x_i$ while $z_a$ is the predicted label according to the attack's purpose that can be targeted or non-targeted. In the white-box attack~\cite{carlini2017towards}, the attacker's knowledge about the $F_{\Theta}(x_i)$ (the architecture and weights) to compute $\Delta$ for a given $x$ is required. In contrast, the attack should be performed without this knowledge under the black-box attack, where repeatedly queries to the model are carried out, and responses are used to guide the crafting of $\Delta$~\cite{papernot2017practical}. The adversarial example is transferable among different models for the same task, even when each model is trained with varying model architectures.

Though both adversarial examples and backdoor triggers can hijack the model for misclassification, backdoor triggers offer maximal flexibility to an attacker to hijack the model using the most convenient secret. Consequently, an attacker has full control over converting the physical scene into a working adversarial input, where backdoor attacks are more robust to physical influences such as viewpoints and lighting~\cite{chou2018sentinet,pasquini2020trembling,wenger2020backdoor,sarkar2020facehack}. The other main difference between adversarial examples and backdoor attacks is the affected stage of ML pipeline, as compared in Fig.~\ref{fig:LifeCycle}. In addition, the adversarial example is usually a crafted perturbation {\it per} input, while the backdoor attack can use the same trigger (pattern patch) to misclassify {\it any} input.

\subsection{Universal Adversarial Patch}

Without loss of generality, the Universal Adversarial Patch (UAP) can be regarded as a special form of adversarial examples. Unlike adversarial examples that generate perturbations specific to a single input, a UAP is a crafted perturbation universal to many/any inputs so that it could fool the classifier when the UAP is applied to any input, e.g., image~\cite{moosavi2017universal}, or text context~\cite{song2020universal}, or voice~\cite{neekhara2019universal}. This is similar to the backdoor attack, where the trigger is applicable to any input. To some extent, the universal adversarial patch (UAP) may be viewed as a `ghost' backdoor as it yields the attack effect similar to backdoor attacks---ghost means it is an intrinsic property of the DL model even that is clean or not backdoored. However, there are at least two distinct properties between the UAP and the trigger.

\begin{enumerate}
    \item A trigger is arbitrary, while a crafted UAP is not arbitrary. Therefore a trigger is under the full control of an attacker, while a UAP depends on the model.
    \item Attacking success rate through the backdoor trigger is usually much higher than the UAP, especially when attackers prefer targeted attacks.
\end{enumerate}


\subsection{Data Poisoning Attack}

Conventionally, a poisoning attack is to degrade the model {\it overall} inference accuracy for clean samples of its primary task~\cite{jagielski2018manipulating}. Poisoning attack is also often called availability attack~\cite{jagielski2020subpopulation,weber2020rab} in a sense that such attack results in lower accuracy of the model, akin to a denial of service attack. In contrast, though a backdoor attack can be realized through data poisoning, a backdoor attack retains the inference accuracy for benign samples of its primary task and only misbehaves in the presence of the secret trigger stealthily. In addition, the conventional poisoning attack is untargeted. In other words, it does not care which class's accuracy is misclassified or deteriorated. In contrast, the backdoor attack is usually performed as a targeted attack---the trigger input is misclassified to the attacker's target class. A unified framework for benchmarking and evaluating a wide range of poison attacks is referred to~\cite{schwarzschild2020just}.


\subsection{Backdoor Attack}

Backdoor attack implants backdoor into the ML model in a way that the backdoored model learns both the attacker-chosen sub-task and the (benign) main task~\cite{gu2017badnets,chen2017targeted}. On the one hand, the backdoored model behaves normally as its clean counterpart model for inputs containing no trigger, making it impossible to distinguish the backdoored model from the clean model by solely checking the test accuracy with the test samples. This is different from the above poisoning attack that deteriorates the overall accuracy of the main task, thus noticeable (and suspicious). On the other hand, the backdoored model is misdirected to perform the attacker's sub-task once the secret trigger is presented in the input, e.g., even regardless of the input's original content.

\section{Background}\label{sec:attackSurfaceVariant}

\begin{figure*}[t]
	\centering
	\includegraphics[trim=0 0 0 0,clip,width=0.75\linewidth]{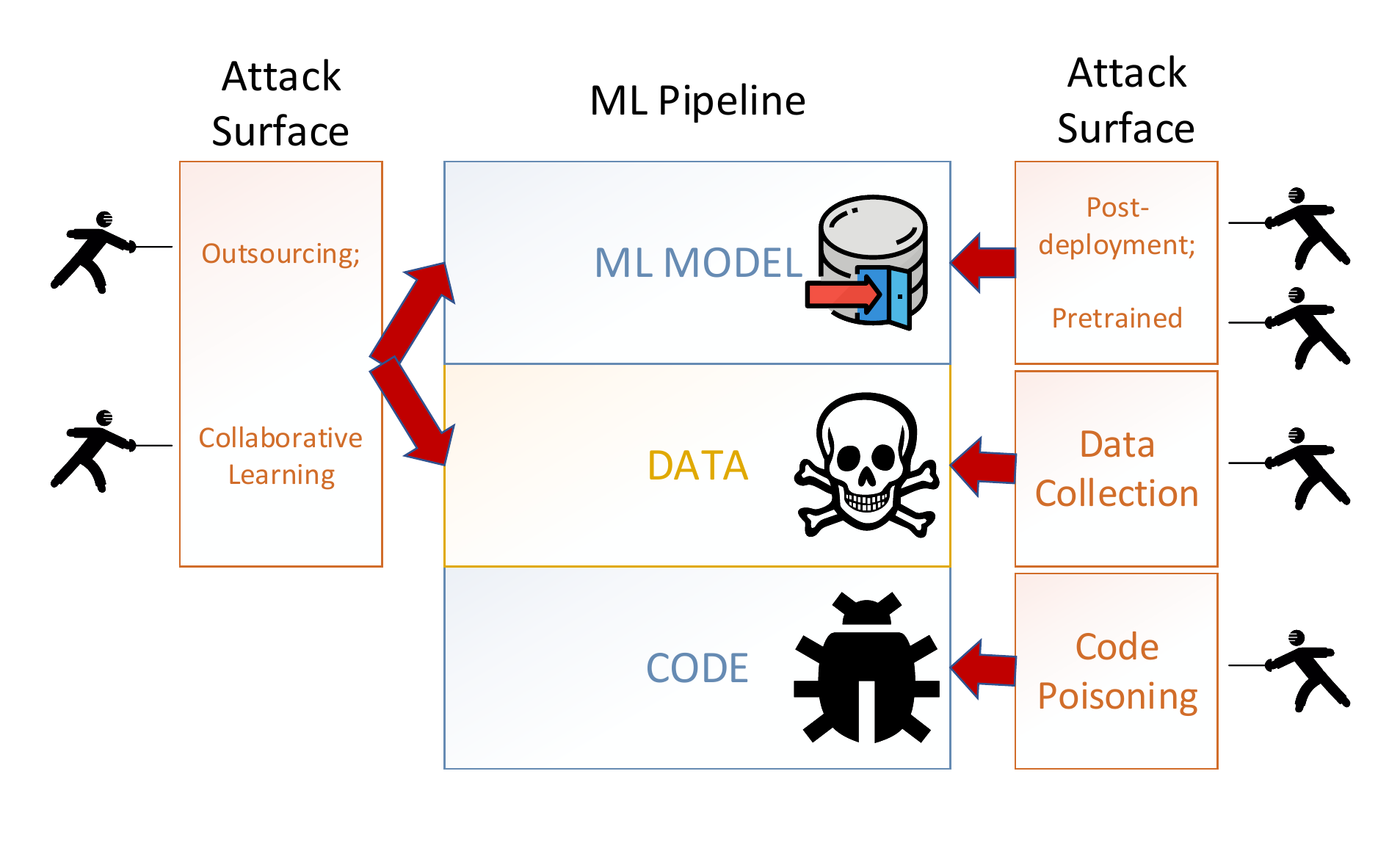}
	\caption{Categorized six backdoor attack surfaces: each attack surface affects one or two stages of the ML pipeline.}
	\label{fig:Threats}
\end{figure*}

In this section, we describe the attack surface of backdoor attacks. Each attack surface could affect different stages of the ML pipeline. We systematically categorize the backdoor attack surface into six groups, as shown in Fig.~\ref{fig:Threats}. We also provide background information about commonly recognized backdoor variants. We then provide some preliminaries such as notations and terms that will be used in the following sections.

\subsection{Attack Surface}~\label{sec:ThreatModel}

\subsubsection{Code Poisoning}

It is common for ML practitioners to employ deep learning frameworks, such as Caffe~\cite{jia2014caffe}, TensorFlow~\cite{abadi2016tensorflow}, and Torch frameworks~\cite{collobert2002torch}, to boost their development. These frameworks are often built over third party software packages that may not be trivial to audit or test. Therefore, a public DL framework could result in vulnerabilities for ML models built upon it. An attacker can exploit the vulnerabilities to launch various attacks ranging from denial-of-service attacks against a DL application to control-flow hijacking attacks that either compromise the system or evades detection~\cite{xiao2018security}. The backdoor attack on DL is already shown to be practical by poisoning the code~\cite{bagdasaryan2020blind}. This attack surface has the weakest attack assumption for an attacker as it has no observation or access to any training data or model architecture, but it would potentially affect the largest number of victims.

\subsubsection{Outsourcing} 

A user will outsource the model training to a third party due to her lack of ML skills or computational resources. In this scenario, the user defines the model architecture, provides training data, and outsources training to  Machine Learning as a Service (MLaaS). As such, a malicious MLaaS provider controls the training phase and backdoors the ML model during the training process.

\subsubsection{Pretrained} 

This attack surface is introduced when a pretrained model or `teacher' model is reused. On the one hand, the attacker can release and advertise a backdoored feature extractor in the image classification to the public, e.g., a model zoo used by a victim for transfer learning~\cite{ji2018model,ji2019programmable}. In natural language processing, the word embedding can act as a feature extractor that may have also been maliciously manipulated~\cite{schuster2020humpty}. Transfer learning is common to train a `student' model with limited data. This is usually the case where data acquisition and labels entail high cost and expertise~\cite{veldanda2020nnoculation}. In addition, transfer learning also reduces computational overhead. On the other hand, the attacker first downloads a popular pretrained benign teacher model, manipulates the original model and retrains it with crafted data (i.e., implanting a backdoor into the model). After that, the attacker redistributes the backdoored model to a model market~\cite{liu2018trojaning,yao2019latent}. It is worth mentioning that the attacker can train a backdoored model from scratch and distribute it to the public.
 

\subsubsection{Data Collection} 

Data collection is usually error-prone and susceptible to untrusted sources~\cite{jagielski2020subpopulation}. If a user collects training data from multiple sources, then data poisoning attacks become a more realistic threat. For example, there are popular and publicly available dataset that rely on volunteers' contribution~\cite{commonvoice,freesound} and/or fetching data from Internet e.g., ImageNet~\cite{deng2009imagenet}. The OpenAI trains a GPT-2 model on all webpages where at least three users on Reddit have interacted~\cite{radford2019language}. As such, some collected data might have been poisoned. They are then released through a data collection bot. When a victim fetches the poisoned data to train a model and even follows a benign training process, the model can still be infected. Specifically, clean-label poisoning attacks~\cite{shafahi2018poison,zhu2019transferable} and image-scaling poisoning attacks~\cite{xiao2019seeing,quiring2020backdooring} lie under such attack surface. Such data poisoning attacks keep consistency between labels and data value, thus bypassing manual or visual inspections.  

\subsubsection{Collaborative Learning} 

This scenario is about collaborative learning represented by distributed learning techniques, e.g., federated learning and split learning~\cite{bagdasaryan2018backdoor,vepakomma2018split,zhou2020privacy}. For example, Google trains word prediction models from data localized on user phones~\cite{hard2018federated}. Collaborative learning or distributed learning is designed to protect privacy leakage on the data owned by clients or participants. During the learning phase, the server has no access to the training data of the participants. This makes collaborative learning vulnerable to various attacks~\cite{bonawitz2019towards}, including the backdoor attack. A jointly learned model can be easily backdoored when a very small number of participants are compromised or controlled by an attacker. Both local data poisoning~\cite{nguyen2020poisoning} and model poisoning~\cite{bagdasaryan2018backdoor,bhagoji2019analyzing,sun2019can} can be carried out by the attacker to implant backdoor into the joint model. We consider the data encryption such as CryptoNet~\cite{gilad2016cryptonets}, SecureML~\cite{mohassel2017secureml} and CryptoNN~\cite{xu2019cryptonn} under this backdoor attack surface, which trains the model over encrypted data in order to protect data privacy. This is, in particular, the case where the data are contributed from different clients, and it is impossible to check whether the data has been poisoned for a backdoor insertion.

\subsubsection{Post-deployment} 

Such a backdoor attack occurs after the ML model has been deployed, particularly during the inference phase~\cite{rakin2020tbt,costales2020live}. Generally, model weights are  tampered~\cite{dumford2018backdooring} by fault-injection (e.g., laser, voltage and rowhammer) attacks~\cite{rakin2020tbt,breier2018practical,hong2019terminal}. Consider a typical attack scenario where an attacker and a user are two processes sharing the same server. The user launches an ML model and loads ML weights into memory. The attacker indirectly flips bits of the weights by triggering the rowhammer bug~\cite{hong2019terminal}, resulting in a downgrade of inference accuracy. 
We note that such an attack cannot be mitigated through offline inspection.


\subsection{Backdoor Variants}~\label{sec:TriggerVariant}
\subsubsection{Class-specific and Class-agnostic}

A backdoor attack is usually targeted because one input is misclassified as the class chosen by an attacker. Backdoor attacks can be categorized into class-agnostic (when the trigger effect is independent of the source classes) and class-specific attacks (when the trigger effect is dependent on the source classes). As for the former, a backdoored model can misclassify input from any class stamped with the trigger into the targeted class. That is, the trigger's presence dominantly determines the attack. As for the latter, a backdoored model can misclassify input from specific classes stamped with the trigger into the targeted class. That is, the trigger's presence, {\it along with the specific class} determines the attack. 

Though we do consider class-specific attacks in this review, we note that majority backdoor attacks and countermeasures are class-agnostic. Therefore, below we list several representative backdoor variants of class-agnostic attacks~\cite{wangneural,gao2019strip}.


\subsubsection{Multiple Triggers to Same Label} {\bf (V1)} In this case, the presence of any trigger among multiple triggers can hijack the backdoored model to the same targeted label.  

\subsubsection{Multiple Triggers to Multiple Labels}  {\bf (V2)} 
In this case, the backdoored model can be hijacked with multiple triggers; however, each trigger targets a different label.


\subsubsection{Trigger Size, Shape and Position } {\bf (V3)} All these trigger settings can be arbitrary to the attacker. For example, the trigger can be any size and in any location. Notably, for the audio and text domains, the shape of the trigger might not be applicable.

\subsubsection{Trigger Transparency } {\bf (V4)} In the vision domain, this occurs when the trigger is blended with the input image. This can also refer to the invisibility of the trigger. In the audio domain, this is related to the trigger amplitude of the audio. In the text domain, this means that the semantic has remained.

We note that most backdoor works have taken V3 and V4 into consideration. Relatively few studies consider other variants.

\subsection{Backdoor Preliminary}


We can formally define the backdoor attack as follows. Given a {\it benign} input $x_i$, on one hand, the prediction $z_a = F_{\Theta_{bd}}(x_i)$ of the backdoored model has a very high probability of being the same as the ground-truth label $z_i$---$z_i$ is $\argmax_{i \in [1,M]} y_i$. In this context, the backdoored model $F_{\Theta_{bd}}$ has equal behavior of the clean model $F_{\Theta_{cl}}$. Herein, $y\in \mathbb{R}^m$ is a probability distribution over the $M$ classes. In particular, the $y_i$ is the probability of the input belonging to $i_{\rm th}$ class (label). On the other hand, given a trigger input $x_i^a = x_i + \Delta$ with the $\Delta$ being the attacker's trigger stamped on a clean input $x_i$, the predicted label will always be the class $z_a$ set by the attacker, even regardless of what the specific input $x_i$ is. In other words, as long as the trigger $x_a$ is present, the backdoored model will classify the input to what the attacker targets. However, for clean inputs, the backdoored model behaves as a benign model without (perceivable) performance deterioration. We note that most backdoor attacks and countermeasures focus on the input-agnostic trigger or class-agnostic trigger. However, some studies focus on the class-specific trigger or class-dependent trigger, which are detailed in \autoref{sec:TriggerVariant}. 

The success of backdoor attack can be generally evaluated by clean data accuracy ({\bf CDA}) and attack success rate ({\bf ASR}), which can be defined as below~\cite{veldanda2020nnoculation}:
\begin{definition}{Clean Data Accuracy ({\bf CDA}):}
The CDA is the proportion of clean test samples containing no trigger that is correctly predicted to their ground-truth classes.
\end{definition}
\begin{definition}{Attack Success Rate ({\bf ASR}):}
The ASR is the proportion of clean test samples with stamped the trigger that is predicted to the attacker targeted classes.
\end{definition}
For successful backdoored model $F_{\Theta_{bd}}$, the CDA should be similar to the clean model $F_{\Theta_{cl}}$, while the ASR is high---backdoored models can achieve an ASR that is close to 100\% usually under \texttt{outsource} attributing to the full control over the training process and data.

\begin{figure}[h]
	\centering
	\includegraphics[trim=0 0 0 0,clip,width=\linewidth]{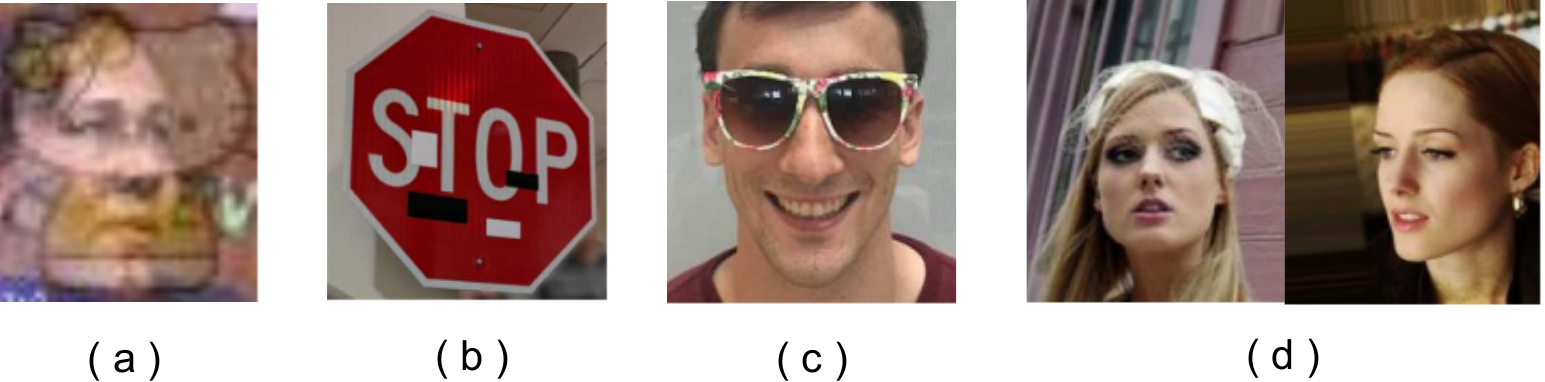}
	\caption{Different means of constructing triggers. (a) An image blended with the Hello Kitty trigger~\cite{chen2017targeted}. (b) Distributed/spread trigger~\cite{eykholt2018robust,guo2019tabor}. (c) Accessory (eye-glass) as trigger~\cite{wenger2020backdoor}. (d) Facial characteristic as trigger: left with arched eyebrows; right with narrowed eyes~\cite{sarkar2020facehack}.}
	\label{fig:TriggerType}
\end{figure}

To ease the readability, we generally provide some terms used frequently in this review.

\begin{enumerate}
    \item The {\bf user} term is exchangeable with the defender, as in most cases, the defender is the end-user of the DNN model. The {\bf attacker} is the one who wants to backdoor the model using any means.
    \item The {\bf clean input} refers to the input that is free of a trigger; it could be the original training sample, validation sample, or test sample. We may alternatively use a clean sample, clean instance, or benign input for easing the description when necessary.
    \item The {\bf trigger input} is the input contains the attacker chosen trigger in order to fire the backdoor. We may alternatively use trigger samples, trigger instance, adversarial input, or poisoned input for easing the description when necessary.
    \item The {\bf target class} or target label refers to the attacker targeted class/label given the trigger input.
    \item The {\bf source class} or source label refers to the class/label within which the attacker chose an input to stamp with the trigger.
    \item The {\bf latent representation} or latent feature, informally, refers to an (often lower-dimensional) representation of high-dimensional data, in particular, the input. Latent representation is the feature from internal layers within a deep neural network in the following descriptions unless otherwise states. As will be witnessed later, a number of countermeasures exploit latent representation to detect backdoor behavior.
    \item {\bf Digital Attack:} Adversarial perturbations are stamped on the digital inputs (e.g., through the modification on the pixels in the digital images).
    \item {\bf Physical Attack:} Adversarial perturbations are physical objects in the physical world where the attacker has no control over the digital input. For example, a physical glass can be a backdoor trigger that will be captured by the face recognition system. The captured digital image is out of the reach of the attacker.
\end{enumerate}

\noindent

\noindent

Notably, physical attacks should be robust against analog-to-digital conversions such as light brightness, noise, angle, and distance variations when capturing images. While digital attacks can use imperceptible perturbations, the physical attack can be readily achieved through backdoor attack~\cite{bhalerao2019luminance,wenger2020backdoor,sarkar2020facehack} using perceptible but inconspicuous triggers such as accessories~\cite{wenger2020backdoor} or even facial expressions~\cite{sarkar2020facehack}, see some trigger examples in Fig.~\ref{fig:TriggerType}. 
Note that most ML models, e.g., self-driving systems, are deployed in an autonomous means, even the perceptible perturbation could be suspicious to humans; the defender could be unaware of it unless the system can throw an alert. The physical attack is more realistic and dangerous~\cite{eykholt2018robust,xiang2019revealing}. 

\section{Backdoor Attacks}\label{sec:backdoorAttack}

We organize the below review mainly according to the attack surface identified in the aforementioned section. We follow the same order except a defer of \texttt{code poisoning} to be last for easing description. For reviewed attacks under each attack surface, a followed highlight is presented. At the end of this section, comparisons and summaries are provided.



\subsection{Outsourcing Attack}


Though earlier neural network backdoor can be dated back to 2013~\cite{geigel2013neural}, the proliferation of backdoor attacks started in 2017 following works from Gu {\it et al.}~\cite{gu2017badnets}, Chen {\it et al.}~\cite{chen2017targeted} and Liu {\it et al.}~\cite{liu2018trojaning}. Gu {\it et al.}~\cite{gu2017badnets} and Chen {\it et al.}~\cite{chen2017targeted} concurrently demonstrate the backdoor attack where the attacker as a third party has access to the training data and the model. Thus, the attacker can manipulate the data or/and the model parameters according to their willingness to insert a backdoor. It is worth mentioning other studies during this early period~\cite{liu2017neural,liu2018sin} with a similar \texttt{outsource} threat model. One common strategy is to stamp the attacker-chosen trigger with (randomly picked) clean samples to form trigger samples, and meanwhile change the labels of these trigger samples to the targeted class. The trained model over normal samples and trigger samples learns to associate the trigger to the target class while maintaining the CDA of clean samples to be similar to a clean model. In other words, the model can efficiently learn the attacker-chosen backdoor sub-task and its main task at the same. Gu {\it et al.}~\cite{gu2017badnets} use a square-like fixed trigger located in the right corner of the digit image of the MNIST data to demonstrate the backdoor attack, with an attack success rate of over 99\% without impacting model performance on benign inputs. In addition, triggers to misdirect traffic sign classifications are studied in~\cite{gu2017badnets}. 


{\it Invisible Trigger.}
There are several studies~\cite{liao2018backdoor,li2019invisible} to make the trigger invisible by humans. Here, the trigger cannot be arbitrary. For example, the trigger can be chosen as a {\it constant change of pixel intensity} added to input (i.e., additive attack)~\cite{liao2018backdoor}. However, it may not be less attractive for performing such an invisible trigger attack if the trigger input's label still needs to be changed---the input content and corresponding label are still inconsistent. In this context, the data poisoning methods under \texttt{data collection} (see Section~\ref{sec:TM3Attack}), such as clean-label poison, tend to be more practical and stealthy, which can easily fool human inspection as well---the input context and label are now consistent. In addition, we regard one strength of the backdoor attack is the arbitrarily chosen trigger to facilitate the attack phase, especially under the physical world. An invisible trigger is hard to survive under, e.g., angle, distance, light variations in the physical world. Such an invisible trigger is unfeasible for some attack scenarios, such as a physical accessory such as glass, earrings, or necklace, to bypass the face recognition system. We note that the invisibility leads the trigger to be more like added noise. To a large extent, the defense against adversary examples can be applied to defense such an invisible backdoor attack. There are several effective adversarial example countermeasures, e.g., feature squeezing~\cite{xu2017feature}, input transformation~\cite{guo2017countering,song2018defense}, noise reduction~\cite{liang2017detecting} that may be applicable to defeat such invisible trigger attacks. 

{\it Dynamic Trigger.} Backdoor attacks usually utilize a static trigger, which is a {\it fixed pattern placed in a fixed location}. The static trigger is trivial to craft for a digital attack but tends to be difficult for physical attack, e.g., the location and angle is hard to be controlled to be consistent when the surveillance camera takes the image. Salem {\it et al.}~\cite{salem2020dynamic} explicitly and systematically study the practicality of dynamic trigger, which can hijack the same label with different trigger patterns that shall share similar latent representation, and locations by exploiting the generative adversarial model to algorithmically generate triggers to facilitate the backdooring. A more straightforward investigation of dynamic triggers is also performed by {\it et al.}~\cite{li2020rethinking}. The dynamic trigger could flexibly craft during the physical attack phase attributing to its constant efficacy under substantial variations. It is shown that dynamic triggers backdoor attack is more powerful, which requires new techniques to defeat since it breaks the static trigger assumption on which most current defenses build~\cite{salem2020dynamic}.

Most backdoor attacks are focusing on classification tasks---mainly the computer vision domain. There exist studies beyond (image) classification.  

{\it Backdoor Reinforcement Learning.} There are examinations on backdoor threats to sequential models~\cite{yang2019design,kiourtitrojdrl}, in particular, reinforcement learning that aims to solve sequential decision problems where an environment provides feedback in the form of a reward to a state. More efforts/considerations are needed to backdoor this sequential model compared with trivially backdooring classification models via training data manipulation. Because a successful attack needs to disrupt the subsequent decisions made by a reinforcement learning policy and not just one isolated decision while maintaining the backdoored model good performance in the absence of trigger~\cite{kiourtitrojdrl}, this enforces the attacker to carefully decide at which timesteps the state is poisoned and the corresponding action and reward are manipulated. Through LSTM based reinforcement learning under \texttt{outsourcing}, Yang {\it et al.}~\cite{yang2019design} has backdoored it to learn an adversarial policy successfully ---the backdoor---to make attacker chosen sequential actions, besides the normal policy to perform the benign model expected actions. For the backdoored sequential model, one distinction is the means of the trigger being presented and backdoor behavior being activated. In previous backdoored classification tasks, the backdoor effect is activated instantly when the trigger is presented and disappears once the trigger is absent. In contrast, Yang {\it et al.}~\cite{yang2019design} demonstrate that the action can be unsynchronized with the presence of the trigger in the backdoored sequential model---the adversarial action can be several steps afterward. In addition, the trigger is only presented a short time of period---e.g., at a specific step, but the backdoor behavior will continue given the trigger being disappeared. These two properties could make the trigger detection more daunting as it becomes infeasible to link the trigger with the adversarial actions even the adversarial action is indeed noticed.

Wang {\it et al.}~\cite{wangstop} examines the backdoor attack on reinforcement learning-based autonomous vehicles (AV) controllers that are used to manage vehicle traffic. By setting the trigger as a specific set of combinations of positions and speeds of vehicles that are collected by various sensors, the AV controller could be maliciously activated to cause a physical crash or congestion when the corresponding triggers appear acting as instructions to accelerate and decelerate. Such backdoor attacks are validated using a general-purpose traffic simulator~\cite{wangstop}.

{\it Code Processing.} Source code processing now also uses a neural network. Ramakrishnan~\cite{ramakrishnan2020backdoors} investigate backdoor attacks on code summarization and classification. For example, the learned model determines whether the source code is {\it safe} or not. To preserve the functionality of the source code, dead-code is employed as a trigger. The authors note that though dead-code may be eliminated by applying a dead-code elimination compiler, it is easy to construct dead-code to bypass it. Under the same research line, Schuster {\it et al.}~\cite{schuster2020you} investigate a backdoor attack on neural network-based code autocompleters, where code autocompletion is nowadays a key feature of modern code editors and integrated development environment (IDE) providing the programmer with a shortlist of {\it likely} completions built upon the code typed so far. The backdoored autocompleter is misdirected to make suggestions for attacker-chosen contexts without significantly changing its suggestions in all other contexts and, therefore, without affecting the overall accuracy. For example, the backdoored autocompleter confidently suggests the electronic codebook (ECB) mode---an old and insecure mode---for encryption for a secure sockets layer (SSL) connection. Both backdoors~\cite{ramakrishnan2020backdoors,schuster2020you} are mainly implemented from training data poisoning, while fine-tuning an existing benign model is also exploited by Schuster {\it et al.}~\cite{schuster2020you}. For both attacks, the triggers are carefully identified or selected in order to strengthen the attack performance.

{\it Backdoor Graph Neural Network.} There are two concurrent backdoor attacks on graph neural network (GNN)~\cite{zhang2020backdoor,xi2020graph}, whereas the work~\cite{xi2020graph} focuses on the \texttt{pretrained} attack surface---description is deferred to following~\autoref{sec:pretrainedAttack}. Considering the discrete and unstructured graph characteristics, the trigger herein is defined as specific subgraph, including both topological structures and descriptive (node and edge) features, which entail a large trigger design spectrum for the adversary~\cite{xi2020graph}. Generally, the GNN can perform two basic graph analytics: node classification that predicts a label for each node in a graph, and graph classification predicts a label for the entire graph. Zhang {\it et al.}~\cite{zhang2020backdoor} study the GNN's vulnerable to backdoor attack and focus on the graph classification task, where a {\it subgraph pattern} is used as a trigger to replace a subgraph within an input graph. The backdoored model is trained through a combination of trigger graphs with i) manipulated targeted label, and ii) clean graphs. High ASR of such a GNN backdoor attack with a small impact on the CDA for clean graphs is validated via three real-world datasets~\cite{zhang2020backdoor}. A certified defense, the particularly randomized subsampling-based defense, is also evaluated, but it is not always effective. Thus, new defenses are required against GNN based backdoor attacks.

{\bf Notes:} \texttt{Outsourcing} backdoor attacks are easiest to perform as the attacker i) has full access to training data and model, and ii) controls the trigger selection and the training process. Therefore, they can carry out data poisoning or/and model poisoning to backdoor the model returned later to the user who outsources the model training. Given control of the training process, it is worth mentioning that the attacker can always take the evasion-of-the-defense objectives into the loss function to adaptively bypass existing countermeasures~\cite{bagdasaryan2018backdoor,costales2020live,tan2019bypassing,bagdasaryan2020blind}.

\subsection{Pretrained Attack}\label{sec:pretrainedAttack}
\begin{figure}[t]
	\centering
	\includegraphics[trim=0 0 0 0,clip,width=\linewidth]{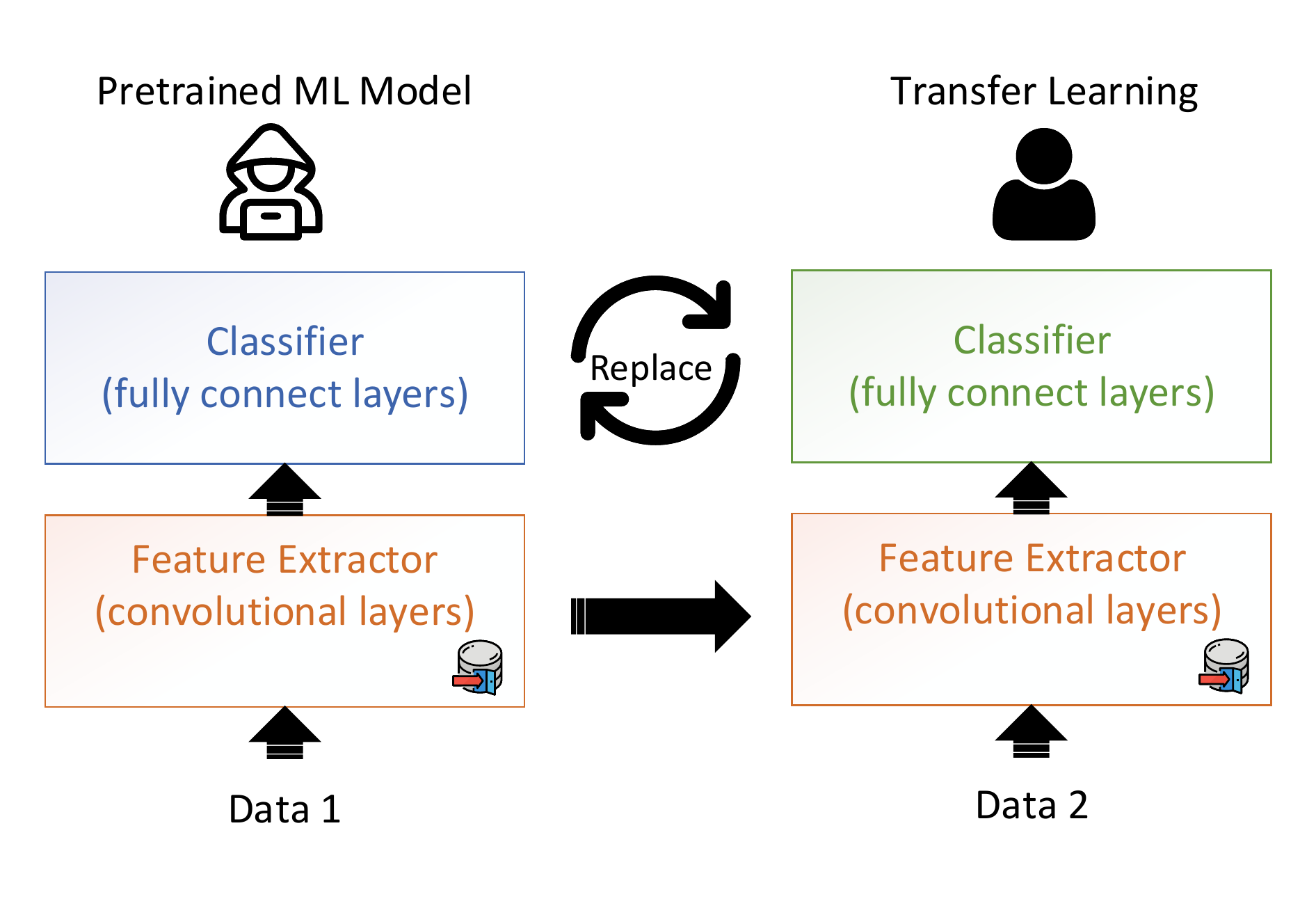}
	\caption{Transfer learning. Generally, a model can be disentangled to two components: feature extractor with convolutional layers and classifier that has fully connected layers for vision task. Usually, the pretrained teacher ML model, e.g., VGG~\cite{simonyan2014very}, is trained over a large-scale dataset, Data 1, such as ImageNet~\cite{deng2009imagenet}, that the user is unable to obtain or/and the training computation is extensive. The user can use the feature extractor of the pretrained ML model to extract general features, gaining an accurate student model given her specific task usually over a small Data 2.}
	\label{fig:Pretrained}
\end{figure}

This attack is usually mounted via a transfer learning scenario, where the user is limited with few data or/and computational resources to train an accurate model. Therefore, the user will use a public or third party pretrained model to extract general features. As illustrated in Fig.~\ref{fig:Pretrained}, here the DNN model, $F_{\Theta}$ can be generally treated as two components, feature extractor $f_{\Theta}$ and task-specific classification layers $g_{\Theta}$. The former is usually performed by convolutional layers to encode domain knowledge that is non-specific to tasks. The later is usually performed by dense layers or fully connected layers. In transfer learning, the users usually replace the dense layers $g_{\Theta}$ but completely retain or only fine-tune feature extractor $f_{\Theta}$. Thus, backdoored feature extractor $f_{\Theta_{\rm bd}}$ entails inherent security risk when it is reused to train an ML system through transfer learning.

{\it Trojan Attack.} Liu~{\it et al.}~\cite{liu2018trojaning} backdoor the published model and then redistribute it. Inserting backdoor in this way eschews overheads of training the entire model from scratch. It does not need to access the published model's original training data because the attacker first carries out reverse engineering to synthesize the training data. Instead of arbitrarily set a trigger, they generate a trigger to maximize the activation of chosen internal neurons in the neural network. This builds a more reliable connection between triggers and internal neurons, thus requiring less training samples to inject backdoors~\cite{yao2019latent}. Once the trigger is generated, they produce two types of inputs: i) the reverse-engineered input and ii) the reverse-engineered input with a trigger, to retrain the model. To expedite retraining, only the neurons between the previously identified neuron and output layers are tuned. In this way, the model retains CDA while achieves a high ASR. Strictly, this attack does not specifically target transfer learning on which the victim trains the student model---to some extent, belongs to \texttt{outsourcing} attack surface. It exploits the published model to expedite the backdoor process and expects the victim to use the redistributed backdoored model {\it directly}. Nonetheless, it is shown if the later victim indeed adopts this backdoored model for transfer learning, the model accuracy on the trigger input will be degraded to some extent---messing up its correct behavior. However, as the attacker does not know the new task's output labels, the trigger input is incapable of misdirecting the new model to a target label.

{\it Model Reuse Attack.} Ji {\it et al.}~\cite{ji2018model} propose the so-called model-reuse attack to backdoor the pretrained model, more precisely, the feature extractor. A surrogate classifier $g_{\Theta}$---could be a simple fully connect layer along with a softmax layer---is used to train a $f_{\Theta_{\rm bd}}$ from its counterpart $f_{\Theta_{\rm cl}}$. Given the attacker-chosen trigger input or, more specifically, the adversarial input, the attacker selectively changes a small fraction of the weights of $f_{\Theta_{\rm cl}}$. 
The change of selected weights maximizes the adversarial input to be misclassified into the target class, while the change has the least degradation on legitimate inputs to be correctly classified. It is shown that the ASR is high, e.g., 90-100\% and the CDA is imperceptibly influenced given that the user deploys $f_{\Theta_{\rm bd}}$ to train an ML system even with fine-tuning on the $f_{\Theta_{\rm bd}}$. 

Model reuse attack~\cite{ji2018model} assumes no knowledge of $g_{\Theta}$ used by the user. The $f_{\Theta_{\rm bd}}$ maybe partially or fully fine-tuned by the user. However, the attacker does need to know the downstream task and a small dataset used by the user.
In contrast to conventional backdoor attacks where any input stamped with the trigger would be misclassified, it appears model-reuse attack is only applicable to limited adversarial inputs ~\cite{ji2018model}. We note that the term `trigger' used in~\cite{ji2018model} has an entirely different meaning and has no generalization. The `trigger' to the backdoored $f_{\Theta_{\rm bd}}$ is one or few attacker chosen {\it specific image inputs}---termed as a single trigger and multiple trigger attacks in~\cite{ji2018model}, respectively. Such an effect can be comparably achieved by adversarial example attacks with the main difference that the attacker arbitrary control over the adversarial example (so-called trigger input(s) herein~\cite{ji2018model}) under model-reuse attack. In addition, as this attack needs to limit the amplitude and number of internal model parameters changed to not degrade the CDA, attack efficacy to DNN model with a relatively small number of parameters or/and binary network work~\cite{qin2020binary}, e.g., those targeting lightweight mobile devices, might be limited.

{\it Programmable Backdoor.} A later work~\cite{ji2019programmable} eschews the assumption on the knowledge of specific task chosen by the user, namely programmable backdoor. Herein, the feature extractor $f_{\Theta_{\rm bd}}$ is trained {\it along with a trigger generator}. The generator, eventually a separate neural network, consists of an encoder and decoder. The encoder encodes a target image into an intermediate feature or latent representation, which is decoded by the decoder into a {\it small} trigger pattern. During the attack phase, this trigger pattern is stamped with a source image--- the trigger pattern is much smaller than the source image---to craft a trigger image. When this trigger image is seen by the student model fine-tuned from $f_{\Theta_{\rm bd}}$ via transfer learning, it will misclassify the trigger image into the target class. The trigger pattern is eventually a smaller representation of the target image chosen from the target class and can be flexibly produced during the attack phase. Therefore, the backdoor is flexibly {\it programmable}. The attacker can choose the target image from any class; such class is unnecessarily known by the attacker when the backdoor is inserted into the pretrained model. Therefore, the number of target class infected is unlimited, which is an advantage over the above model reuse attack~\cite{ji2018model}. Similar to~\cite{ji2018model}, its attack efficacy is dependent on the size of the $f_{\Theta_{\rm cl}}$---redundancy information. The CDA drop could be noticeable, and ASR is lower when the size of the target model $f_{\Theta_{\rm cl}}$ is small, e.g., MobileNet~\cite{howard2017mobilenets}. In addition, fine-tuning the entire $f_{\Theta_{\rm bd}}$ rather than solely the fully connected layers $g_{\Theta}$ can greatly reduce the ASR, though the ASR cannot be fundamentally eliminated.

{\it Latent Backdoor.} Yao {\it et al.}~\cite{yao2019latent} propose the latent backdoor attack to infect the pretrained teacher model. Generally, a latent backdoor is `incomplete' backdoor injected into the teacher model---the attacker target label $z_a$ does not exist in the teacher model yet. However, if any student models include $z_a$, the transfer learning process automatically completes the backdoor and makes it active. Recall that the attacker has no control over the user transfer learning process. Then the attacker can attack the student model with the trigger determined when backdooring the student model. To be precise, this attack follows the below steps. Firstly, the attacker chooses a clean teacher model. she also determines the future target class $z_a$, which is {\it not} currently within the teacher task. However, the attacker does need to collect source samples of $z_a$, e.g., from the public. Here, we take an example, where VGG16 is a clean teacher model, and Musk that is not a recognizable face of VGG16 is the future target label $z_a$. Secondly, the attacker retrains teacher model to i) include $z_a$ as a classification label, ii) injects latent backdoor related to $z_a$, then iii) replaces the classification layer to remove output $z_a$, and iv) preferably further fine-tunes on the last layer. In this way, the backdoored model has the same benign functionality with the clean model. Thirdly, a victim downloads the backdoored teacher model and applies transfer learning to customize the user's downstream task, which includes $z_a$ as one of the classes. As a result, the latent backdoor will be activated as a live backdoor in the student model. For example, Tesla builds its facial recognition task building upon the infected VGG16 and adds Musk as one class. During the attack, any image with the trigger will be misclassified to Musk by the backdoor propagated student model. The key in this latent backdoor attack is to associate trigger to the intermediate representation within the internal layer, e.g., 14th layer out of total 16 layers in the teacher model, instead of associating with the final output label adopted by conventional backdoor attacks. To amplify the ASR, the trigger is not randomly chosen but optimized through an algorithm. The latent backdoor has the advantage of stealthiness. However, it only works when the downstream task contains the attacker speculated label $z_a$. It easily bypasses NeuralCleanse~\cite{wangneural} defense that iteratively goes through all output labels to discover triggers because the teacher model does not have the infected label $z_a$ yet. However, the ASR drops to almost zero if the user fine-tunes earlier, e.g., 10th layer before the layer that associates the intermediate trigger representation, e.g., 14th. As a matter of fact, the latter fine-tune strategy is not uncommonly used in practice.


{\it Appending Backdoor.} Tang {\it et al.} propose a different model-agnostic backdoor attack without access to neither training data nor training the targeted model~\cite{tang2020embarrassingly}. It is also with the advantage of an unlimited number of target classes as programmable backdoor~\cite{ji2019programmable}. Overall, it achieves so by {\it appending} or merging another separate backdoored (small) neural network with the target model. Therefore, it does not tamper the parameters of the target model. As a trade-off, this work {\it has to change the model architecture}. In practice, architecture change is not stealthy; thus, it is easily detectable by simply checking the model architecture specifications.

{\it GNN Backdoor.} Concurrent to this work~\cite{zhang2020backdoor} that evaluates the graph neural network (GNN) vulnerability to backdoor under~\texttt{outsource} attack surface, Xi {\it et al.}~\cite{xi2020graph} investigate the GNN vulnerability to backdoor under \texttt{pretrained} attack surface. Here, it is assumed~\cite{xi2020graph} the attacker has access to the downstream task and a set of data samples. However, similar to~\cite{ji2019programmable,yao2019latent}, it does not need knowledge of downstream classifier $g_{\Theta}$ and fine-tuning strategies applied by the user. The attacker can train the pretrained GNN model $f_{\Theta_{\rm bd}}$ from scratch or a published model. Here, the graph trigger is interactively optimized along with the training of $f_{\Theta_{\rm bd}}$ to increase the ASR and stealthiness. The later is achieved by adaptively tailoring graph trigger according to the given graph and mixing it within the graph to replace a subgraph. The subgraph within the given graph that is found by a {\it mixing function} is most similar to the graph trigger. Based on five public dataset validations~\cite{xi2020graph}, it is shown the ASR could be up to 100\% and with comparable CDA to clean graph model. It is also shown that current Neural Cleanse~\cite{wangneural} fails to defeat the GNN backdoor.

{\bf Notes:} \texttt{Pretrained} backdoor attacks have broad victim impacts as employing a pretrained teacher model to customize the user's downstream task is now a good practice to expedite the DNN model development with high accuracy even with limited training data or/and computational resources. Besides vision tasks, such attacks have also been applied to natural language processing~\cite{kurita2020weight,schuster2020humpty}. However, the attacker has no control of the downstream tasks and the transfer learning strategies adopted by the user. The ASR is usually not high as \texttt{outsourcing} backdoor attacks. As for the latent backdoor~\cite{yao2019latent}, the ASR can be easily disrupted. It is worth to mention that the \texttt{pretrained} backdoor attack, more or less, assumes specific knowledge of the downstream tasks (may be speculated) and a small set of data for the downstream tasks (may be collected from the public).

\subsection{Data Collection Attack}~\label{sec:TM3Attack}
Data collection attack succeeds to poison data, and consequently backdoor the model without inside access. This needs the poisoned data to be stealthier, in particular, to fool human inspection. However, the attacker cannot control and manipulate the training process or the final model.

\begin{figure}[h]
	\centering
	\includegraphics[trim=0 0 0 0,clip,width=\linewidth]{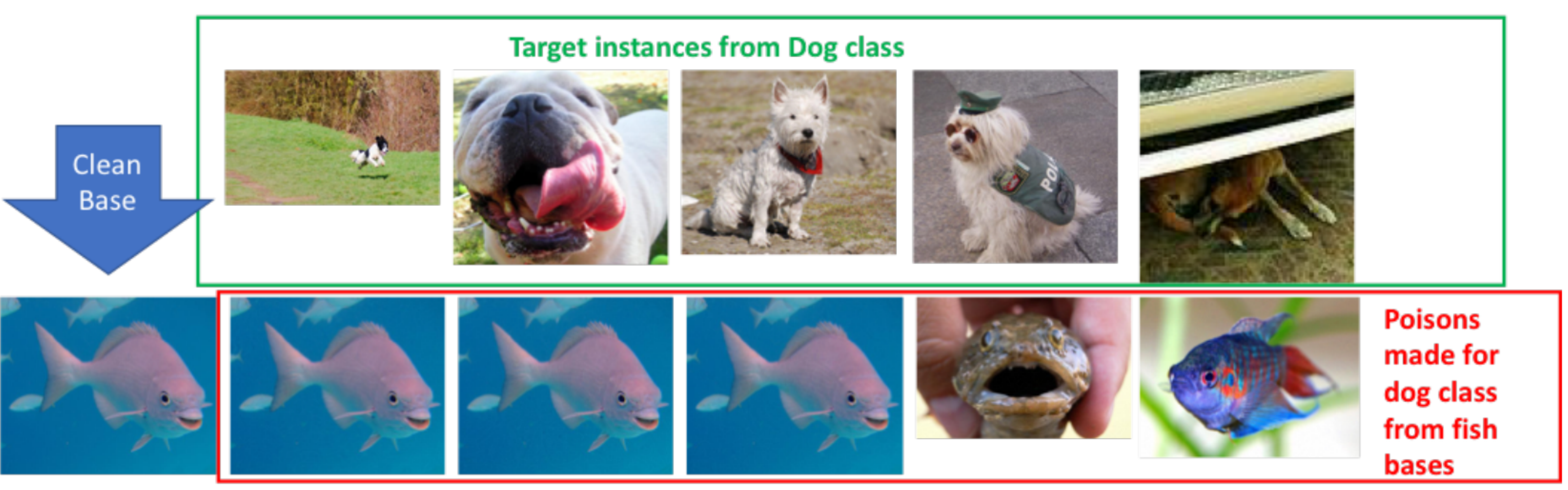}
	\caption{Clean-label attack~\cite{shafahi2018poison}.}
	\label{fig:cleanLabel}
\end{figure}

{\it Clean-Label Attack.} Clean-label attack preserves the label of poisoned data, and the tampered image still looks like a benign sample~\cite{shafahi2018poison,zhu2019transferable}. As shown in Fig~\ref{fig:cleanLabel} as an example, the fish image has been poisoned, where its latent feature, however, represents a dog. Here, the fish is the base/source instance, and the dog is the target instance. Domain experts will label such a poisoned sample as fish. Thus, the attacker needs no control of the labeling process. After the model is trained over poisoned (e.g., fish samples) and clean samples. Such a backdoored model would misclassify targeted instance, e.g., dog into the source instance, e.g., fish. Notably, the targeted instance {\it does not require perturbation during inference}---no trigger is used. Clean-label attack exploits feature collision. To be precise, the crafted poison fish still appears like the source fish instance in the input space or pixel level (e.g., the visual appearance), but it is close to the targeted dog instance dog in latent feature space when features are more abstract such as in the penultimate layer. The clean-label attack demonstrated in~\cite{shafahi2018poison} is the class-specific, and the targeted success rate is low, e.g., 60\%, in end-to-end attack (the victim model is trained from scratch rather than pretrained model) for a 10-class classification setting. Clean-label attack in~\cite{shafahi2018poison} is performed under a white-box setting, where the attacker has complete knowledge of the victim model. Such constraint is eschewed in later improvements that work under black-box settings and dramatically increase attack success rate~\cite{zhu2019transferable,huang2020metapoison}. Instead of feature collision attack, the efficacy of later works are realized via exploiting convex polytope attack~\cite{zhu2019transferable} and bi-level optimization problem~\cite{huang2020metapoison}, respectively, to poison clean-label samples. In addition to image classification, clean-label attacks have been applied to backdoor video recognition models~\cite{zhao2020clean}. For the clean-label attacks, we note the triggers are a particular set of (testing) image samples---not a universal specific pattern. In other words, the backdoor attack is only possible for {\it one or a small set of inputs}---note model-reuse attack~\cite{ji2018model} under \texttt{pretrained} achieves a similar backdoor effect.

Interestingly, Saha {\it et al.}~\cite{saha2019hidden} exploits the clean-label attack in a robust manner, namely hidden trigger backdoor. Here, the trigger used in the poisoning phase or the poisoning trigger is invisible to evade human inspections. However, the trigger used in inference is visible to increase the attack robustness that is important for a physical attack. The key is that the trigger used in the poison phase shares the same latent representation with the trigger used in the inference phase. Therefore, even the inference trigger is not explicitly seen during the training phase, it can still activate the backdoor. Such a hidden backdoor attack appears to be more practical and robust compared with~\cite{liao2018backdoor,li2019invisible}, because a visible inference trigger is only revealed when an attack is presented but hidden during data poisoning via an invisible poisoning trigger. These clean-label attacks assume knowledge of the model to be used by the victim, e.g., to perform transfer learning~\cite{saha2019hidden}. The reason is that the attacker needs to know the latent representation.

Severi {\it et al.}\cite{severi2020exploring} note that the training pipeline of many security vendors naturally enables backdoor injection for security-related classifiers, in particular,  malware classifier. Because these vendors or companies heavily rely on crowd-sourced threat feeds (records) to create a large, diverse stream of data to train the classifier due to the difficulty of exclusively creating in-house data sources. The issue of labeling malware into families, or determining if an unknown binary is or is not malware, is labor-intensive and requires significant domain knowledge and training~\cite{raff2020malwaresurvey}. For example, an expert analyst can often take around 10 hours to characterize a malicious binary. This is in contrast to many current machine learning domains, like image classification, where individuals can often label with no specialized expertise and with acceptable time. This threat feeds, or data sources are built upon user-submitted binaries, which can be readily abused by attackers for backdoor attacks~\cite{severi2020exploring}. Severi {\it et al.} develop a clean-label attack to backdoor malware classifiers of both gradients boosting trees and a neural network model, even when the attacker has black-box access to the user deployed models by only poising a small fraction, e.g., 2\%, of malware. The poisoned goodware/malware is still recognized as goodware/malware by anti-virus engine.

{\it Targeted Class Data Poisoning.} Barni {\it et al.}~\cite{barni2019new} recognize that conventional backdoor introduced by data poisoning, which {\it randomly} i) picks up training samples across different source classes, ii) stamps triggers, and iii) {\it modifies their labels to the targeted class}. Noting, in clean-label attack~\cite{shafahi2018poison,zhu2019transferable,saha2019hidden}, the label is not modified, but the poisoned source images could be from different classes. Therefore, Barni {\it et al.}~\cite{barni2019new} slightly changes the way of picking up data for poisoning. It is shown that it is feasible to only stamp triggers on the data samples belonging to the targeted class---thus, the label is consistent with the content---to backdoor the model to misclassify any input to the targeted label. This is because the model can learn the association between the trigger and the targeted class. Together with an invisible, semantically indistinguishable trigger to humans, such poisoned data can evade the inspection from humans. This data poisoning strategy is used in~\cite{liu2020reflection} to implant backdoor with a delicate crafted natural trigger, namely refection, that is a common natural phenomenon wherever there are glasses or smooth surfaces. One trade-off of this targeted class data poisoning is the fraction of poisoned data of the targeted class needs to be high~\cite{turner2019label}.

Turner {\it et al}~\cite{turner2019label} relaxes the above limitation~\cite{barni2019new}. One observation is that poisoned inputs can be modified to make the model harder to classify to its ground-truth label. Since these inputs will be harder to learn, the model will be {\it forced to learn a strong association between the trigger to the label}. In this way, the poison ratio will be reduced. Two label-consistent data poisoning methods: GAN-based and adversarial perturbations are proposed to inject backdoor. The GAN method interpolates poisoned inputs towards a source class in the {\it latent space}---similar to feature collision, while the poisoned input visually consistent to its label that is indeed the target label. For the adversarial perturbations, it exploits the method of crafting adversarial examples. Before adding triggers to an input, perturbation is firstly made to this input to make the model harder to learn it. Then the (invisible) trigger is added to the adversarial perturbed input to form a poisoned input. To be effective for the ASR, the attacker needs to have full knowledge of the model architecture and later training procedure~\cite{turner2019label}.

\begin{figure}[t]
	\centering
	\includegraphics[trim=0 0 0 0,clip,width=\linewidth]{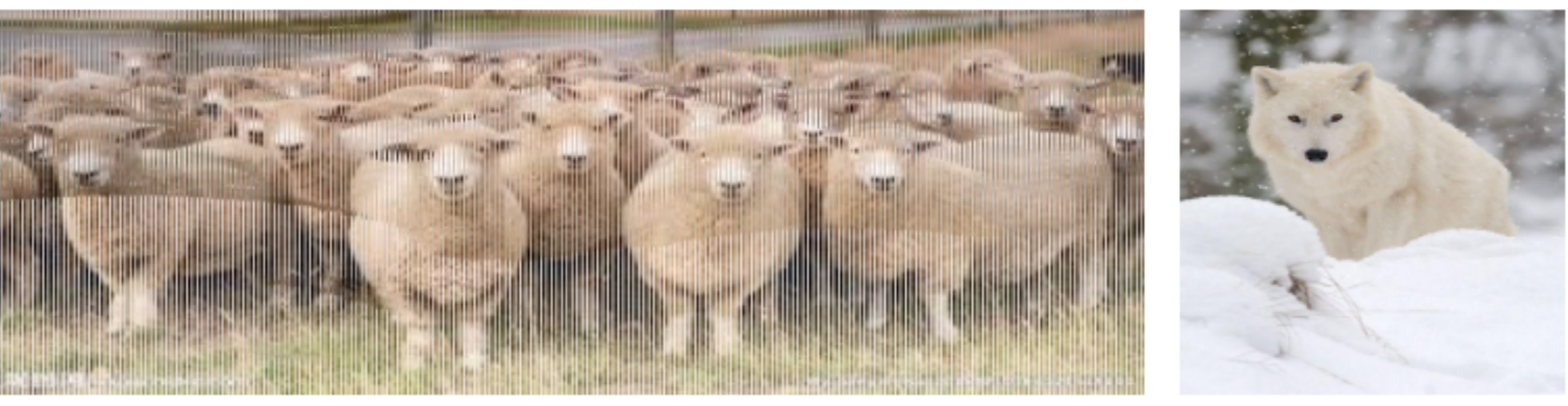}
	\caption{Image-scaling attack~\cite{xiao2019seeing}. The left (labeled) sheep is what humans see, and the right wolf is the actual input fed into the DNN model {\it after scaling} to a fixed size (e.g., $224\times 224\times 3$) that is a commonly used resize process in deep learning. Notably, the perturbation could be more imperceptible; here, the perceptible perturbation is used for demonstration purposes.}
	\label{fig:ImageScaling}
\end{figure}

{\it Image-Scaling Attack.} Image-scaling attack or scaling-camouflage attack~\cite{xiao2019seeing} has no knowledge of training data, model, and parameters, which affects all applications that utilize scaling algorithms for resizing input images. In terms of deep learning, an image-scaling attack abuses the scaling function in the image preprocessing pipeline that is popular in deep learning frameworks such as Caffe, TensorFlow, and PyTorch. It is common the input size to the model is fixed, e.g., $224\times 224 \times 3$ with 224 is height and width, and 3 is the color channel. However, the original image is usually much larger than this input size and has to be resized via downsampling operation. One attack example is illustrated in Fig.~\ref{fig:ImageScaling}. The `wolf' image is embedded/disguised {\it delicately} into the `sheep' image by the attacker by abusing the resize() function. When the down-sampling filter resizes the attack image (tampered `sheep' image), the `sheep' pixels are discarded, and the `wolf' image is presented that is seen by the model. This abuses an inconsistent understanding of the same image between humans (tampered `sheep' image) and machines (`wolf' image). Therefore, scaling camouflage can be effectively and stealthily exploited to perform backdoor attacks under a black-box setting; it does not require to have control over the labeling process~\cite{quiring2020backdooring}. Generally speaking, the {\it trigger images} can be disguised into images of the target class to poison training data by performing image-scaling attacks, thus, backdooring the trained model. In the inference phase, the trigger will easily activate the backdoored model {\it now without using a scaling attack}.

{\bf Notes.} We can observe that the {\it data collection} attack is also with a wide range of victims since it is a common strategy to gain training data from the public. A trust data source is hard to be ensured. For the poisoned data, it would be hard to be distinguished when they undergo manual, or visual inspection since the content is always consistent with the label. As a consequence, not only the end-to-end trained model but also transfer learning could be infected. Feature collision is a common means of crafting label-consistent poisonous inputs to inject backdoors. However, in some cases, some knowledge of the infected model architecture is required to determine a proper latent representation.

\subsection{Collaborative Learning Attack}
\begin{figure}[h]
	\centering
	\includegraphics[trim=0 0 0 0,clip,width=\linewidth]{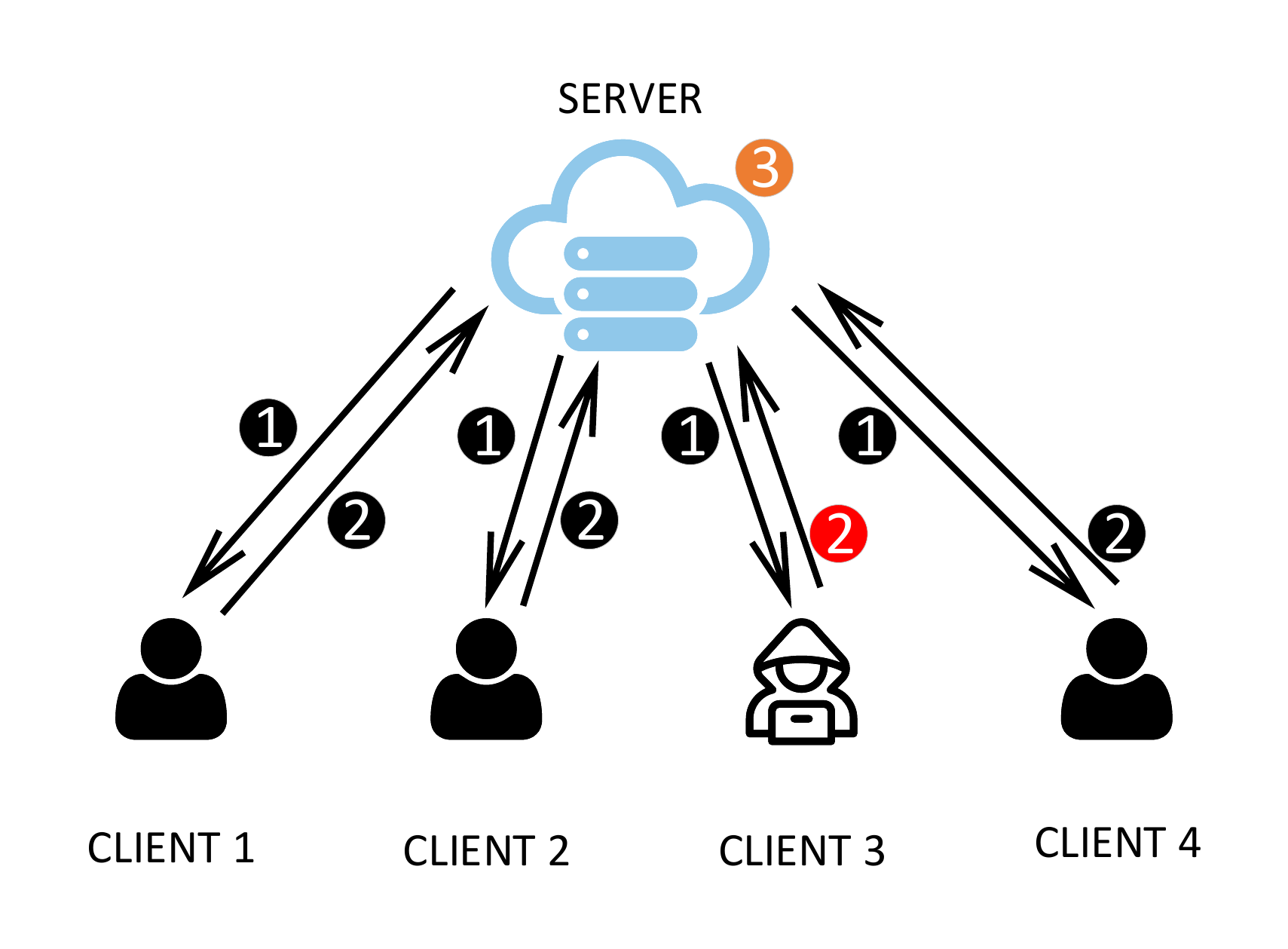}
	\caption{Illustration for federated learning with four clients.}
	\label{fig:FL}
\end{figure}

Collaborative learning aims to learn a joint/global model with distributed clients but does not access data on the participant side. Federated learning is the most popular collaborative learning technique nowadays~\cite{konevcny2016federated,gao2020end}. Federated learning interacts with clients many rounds. As shown in~\ref{fig:FL}, in each round, \textcircled{1} the server sends a joint model---it is randomly initialized in the first round---to all clients, each client trains this model over her local data for one or more local epochs---now the updated model is termed as local model. \textcircled{2} Then the server collects local models from a fraction of client---could be all clients---and \textcircled{3} applies an aggregation algorithm, e.g., \textsf{average}, to update the joint model. This process continues until the model converges or reaches the predefined criterion. In this way, the joint model is trained, while local data never leaves the client's hand. The collaborative learning can significantly protect sensitive data privacy without seeing it, it is, however, vulnerable to backdoor attacks.

Bagdasaryan~{\it et al.}~\cite{bagdasaryan2018backdoor} apply {\it model replacement} to implant backdoor into the joint model. Malicious client, e.g., client 3, controlled by the attacker uploads modified local model, e.g., in step \textcolor{red}{\textcircled{2}} as shown in Fig.~\ref{fig:FL} that are manipulated via optimization, to the server. As a result, the the joint model will be backdoored. This model replacement strategy is also referred to as model poisoning~\cite{bagdasaryan2018backdoor,bhagoji2019analyzing,sun2019can}. It is shown that the joint model is with 100\% ASR {\it immediately after} the backdoor being inserted even if a {\it single client} is selected to contribute for the joint model update in {\it only one round} (single-shot attack). The ASR does diminish as the joint model continues to learn. By controlling no more than 1\% clients, the joint model is with on par accuracy with the clean model and now cannot prevent the backdoor being unlearned afterward~\cite{bagdasaryan2018backdoor}. The backdoor attack on federated learning is hugely challenging to defeat as the data access is not allowed in principle. Byzantine-resilient aggregation strategies are ineffective against such attack~\cite{bhagoji2019analyzing}. Even worse, when secure aggregation is adopted to enhance privacy further, the defense would become more difficult~\cite{bagdasaryan2018backdoor,sun2019can}. When the backdoor defense requires no training data, e.g., DeepInspect~\cite{chen2019deepinspect}, it inverts the model to extract the training data, which unfortunately potentially violates the data privacy preservation purpose of adopting federated learning. Though Sun {\it et al.}~\cite{sun2019can} demonstrate norm clipping and weak differential privacy may limit the attack to some extent, they tend to be easily evaded with slight increased attention and efforts by the attacker, especially when the adaptive attack is considered~\cite{bagdasaryan2018backdoor,bhagoji2019analyzing}. 

Xie {\it et al.}~\cite{xie2019dba} investigate distributed backdoor attack (DBA) on federated learning by fully taking advantage of the distribute characteristics. Instead of assigning all attacker-controlled clients with the same global trigger to poison their local models, the DBA decomposes the global trigger into different (non-overlap) local triggers. Each malicious client poisons her local model with her own local trigger. As an example with four stickers that each sticker acts as one local trigger is shown in Fig.~\ref{fig:TriggerType} (b). The joint model will not exhibit a backdoor effect if only one local trigger presents but will show the backdoor effect for the combined local triggers---the global trigger during the attack phase. It is shown that the DBA is with better ASR and more robust against robust federated learning aggregation methods than previous backdoor attacks on federated learning~\cite{xie2019dba}. It is worth mentioning that both DBA and conventional (global) trigger attacks achieve the same attack purpose---using the global trigger to misdirect the backdoored joint model.

Chen {\it et al.}~\cite{chen2020backdoormeta} study backdoor attack on the federated meta-learning. Meta-learning, also known as ``learning to learn,'' aims to train a model that can learn new skills or adapt to new environments rapidly with a few training examples. Therefore, federated meta-learning allows clients to have a joint model to be customized to their specific tasks, even the data distribution among clients vary greatly. This means the final model is built upon the joint model with further updates with his specific local data, e.g., through fine-tuning. As one prerequisite, the attacker needs to know the target class of the victim. The attack surface of federated meta-learning is alike \texttt{pre-trained} but with a weaker control on the joint model indirectly---attacker has full control over the pre-trained model under \texttt{pre-trained}. As a result, it is shown that properly adopt the (backdoored) joint model to a client-specific task can, to a large extension, reduce the backdoor effect~\cite{chen2020backdoormeta}. Specifically, the authors propose a sanitizing fine-tuning process building upon matching networks~\cite{vinyals2016matching} as a countermeasure, where the class of input is predicted from the cosine similarity of its weighted features with a support set of examples for each class. It is shown the ASR, in this regard, is reduced to only 20\% validated through Omniglot and mini-ImageNet datasets.

{\bf Notes.} On the one hand, collaborative learning becomes increasingly used. On the other hand, it is inherently vulnerable to backdoor attacks because it is hard to control the data and the local model submitted by malicious participants. Due to the privacy protection, the training data is not allowed to be accessed by the defender, in particular, the server or the model aggregator. Such restriction is also applicable when training a model over encrypted data such as CryptoNet~\cite{gilad2016cryptonets}, SecureML~\cite{mohassel2017secureml} and CryptoNN~\cite{xu2019cryptonn}. This makes the defense to be extremely challenging. Because most defense mechanisms indeed need a (small) set of hold-out validation samples to assist the backdoor detection. It is noticed that DeepInspect~\cite{chen2019deepinspect} does not need to access training data. Unfortunately, it, however, inverts the model to reverse engineer the training data, which could partially violates the data privacy preservation purpose of adopting collaborative learning. 

\begin{table*}
	\centering 
	\caption{Backdoor attack summary under categorized attack surfaces. The 3rd to 7th columns qualitatively compare the attacker's capabilities under the corresponding attack surface.}
			\resizebox{1.0\textwidth}{!}{
	\begin{tabular}{c | c | c | c | c | c | c | c | c} %
		\toprule 
		\toprule 
				
		\begin{tabular}{@{}c@{}} Attack \\ Surface \end{tabular} & \begin{tabular}{@{}c@{}} Backdoor Attacks  \end{tabular} & \begin{tabular}{@{}c@{}} Access Model \\ Architecture \end{tabular}  & \begin{tabular}{@{}c@{}} Access Model \\ Parameters \end{tabular} & \begin{tabular}{@{}c@{}} Access \\ Training Data \end{tabular} & \begin{tabular}{@{}c@{}} Trigger \\ controllability \end{tabular} & ASR & \begin{tabular}{@{}c@{}} Potential \\ Countermeasure $^1$ \end{tabular}\\ 
		\midrule
	    \begin{tabular}{@{}c@{}} \texttt{Code} \\ \texttt{Poisoning} \end{tabular}   &  \begin{tabular}{@{}c@{}} \textcolor{blue}{\cite{xiao2018security}}\cite{bagdasaryan2020blind} \end{tabular} & Black-Box & \Circle &  \Circle & \LEFTcircle & High & \begin{tabular}{@{}c@{}} Offline Model Inspection \\ Online Model Inspection \\ Online Data Inspection \end{tabular} \\ \hline	
		
		\texttt{Outsourcing} & \begin{tabular}{@{}c@{}}
		Image~\textcolor{blue}{\cite{liu2017neural,gu2017badnets,liu2018sin,kwon2020multi,truong2020systematic}}~\cite{chen2017targeted}; \\ Text~\textcolor{blue}{\cite{dai2019backdoor}}\cite{sun2020neuralbackdoor,gao2019design,chen2020badnl}; \\ Audio~\cite{gao2019design,kong2019adversarial}; \\Video~\textcolor{blue}{\cite{bhalerao2019luminance}}; \\ Reinforcement Learning~\cite{yang2019design,wangstop}\textcolor{blue}{\cite{kiourtitrojdrl}} \\ (AI GO~\cite{go2020poison}); \\ Code processing~\cite{ramakrishnan2020backdoors,schuster2020you}; \\ Dynamic trigger~\cite{salem2020dynamic} \\ Adaptive Attack~\textcolor{blue}{\cite{tan2019bypassing}}; \\ Deep Generative Model~\cite{ding2019trojan}; \\
		Graph Model~\cite{zhang2020backdoor}\end{tabular} & White-Box & \CIRCLE &  \CIRCLE & \CIRCLE & Very High & \begin{tabular}{@{}c@{}} Blind Model Removal \\ Offline Model Inspection \\ Online Model Inspection \\ Online Data Inspection \end{tabular} \\ \hline
		
		\begin{tabular}{@{}c@{}} \texttt{Pretrained} \end{tabular} & \begin{tabular}{@{}c@{}} \textcolor{blue}{\cite{liu2018trojaning,gu2017badnets}} \\ Word Embedding~\textcolor{blue}{\cite{schuster2020humpty}}; \\
		NLP tasks~\cite{kurita2020weight};\\ Model-reuse~\textcolor{blue}{\cite{ji2018model}}; \\ Programmable backdoor~\cite{ji2019programmable}; \\Latent Backdoor~\cite{yao2019latent}; \\ Model-agnostic via appending~\textcolor{blue}{\cite{tang2020embarrassingly}}; \\ Graph Model~\cite{xi2020graph}\end{tabular} &  Grey-Box & \LEFTcircle &  \LEFTcircle & \LEFTcircle & Medium & \begin{tabular}{@{}c@{}} Blind Model Removal \\ Offline Model Inspection \\ Online Model Inspection \\ Online Data Inspection \end{tabular}\\
		\hline
		
		\begin{tabular}{@{}c@{}} \texttt{Data} \\ \texttt{Collection} \end{tabular} &  \begin{tabular}{@{}c@{}} Clean-Label Attack~\textcolor{blue}{~\cite{shafahi2018poison,zhu2019transferable,saha2019hidden}}~\cite{liu2020reflection}, \\ (video~\textcolor{blue}{\cite{zhao2020clean,bhalerao2019luminance}}), \\ (malware classification~\textcolor{blue}{\cite{severi2020exploring}}); \\ Targeted Class Data Poisoning~\textcolor{blue}{\cite{barni2019new,turner2019label}}; \\ Image-Scaling Attack~\textcolor{blue}{\cite{xiao2019seeing,quiring2020backdooring}}; \\ Biometric Template Update~\cite{lovisotto2019biometric}; \\ Wireless Signal Classification~\cite{davaslioglu2019trojan} \end{tabular} & Grey-Box & \LEFTcircle &  \LEFTcircle & \LEFTcircle & Medium &\begin{tabular}{@{}c@{}} \\ Offline Data Inspection \\ Online Model Inspection \\ Online Data Inspection \end{tabular} \\ \hline		
		
	    \begin{tabular}{@{}c@{}} \texttt{Collaborative} \\ \texttt{Learning} \end{tabular}   &  \begin{tabular}{@{}c@{}} Federated learning~\textcolor{blue}{\cite{bagdasaryan2018backdoor,bhagoji2019analyzing}}, \cite{sun2019can}, \\ (IoT application\textcolor{blue}{~\cite{nguyen2020poisoning}}); \\
	    Federated learning with \\ distributed backdoor~\textcolor{blue}{\cite{xie2019dba}}; \\ Federated meta-learning~\cite{chen2020backdoormeta}; \\ feature-partitioned \\ collaborative learning~\cite{liu2020backdoor} \end{tabular} & White-Box & \CIRCLE &  \CIRCLE & \CIRCLE & High & \begin{tabular}{@{}c@{}} Offline Model Inspection $^2$ \end{tabular} \\ \hline
	    
	    \texttt{Post-deployment} &  \begin{tabular}{@{}c@{}} \cite{dumford2018backdooring}\textcolor{blue}{\cite{rakin2020tbt,costales2020live}} \\ Application Switch~\cite{guo2020trojannet} \end{tabular}   & White-Box & \CIRCLE &  \CIRCLE & \LEFTcircle & Medium & \begin{tabular}{@{}c@{}} Online Model Inspection \\ Online Data Inspection \end{tabular} \\ \hline
		\bottomrule
	\end{tabular}}
	  \begin{tablenotes}
      \small
     \item \CIRCLE: Applicable or Necessary; \CIRCLE: Inapplicable or Unnecessary; \LEFTcircle: Partially Applicable or Necessary.
      \item $^1$ The detailed countermeasure methods are specified in \autoref{sec:backdoorCountermeasure} and summarised in Table~\ref{tab:countermeasure}. We consider they are {\it potential}, because it is recognized that there is no single one-against-all countermeasure and they could be bypassed, especially by adaptive attackers.
      \item $^2$ This kind of backdoor attack is very challenge to defeat as the user, e.g., the model aggregator, may not even be able to access any testing data. Rare defense requires no access to training data~\cite{chen2019deepinspect}.
      
    \end{tablenotes}
	\label{tab:ThreatModel}
\end{table*}

\subsection{Post-Deployment Attack} 

This attack tampers the model during the post-deployment phase and thus does not rely on data poisoning to insert backdoors. However, it inevitably needs model access, which can be realized by firstly insert a malicious software program. Stealthier, it is feasible considering the rising of side-channel attacks, more specifically, utilizing the row-hammer attack~\cite{kim2014flipping}. The row-hammer attack allows an attacker to tamper the memory content if the victim (CNN model) is co-located at the same physical machine, typical in today's cloud service. For example, many tenants are sharing the same physical machine with separation from virtual machine~\cite{zhang2020detecting}.

{\it Weight Tamper.} As an early work, Dumford {\it et al.}~\cite{dumford2018backdooring} examine the feasibility of backdooring a face recognition system through solely perturbing weights that are resided in specific layers---a similar study is demonstrated in~\cite{costales2020live}. Both work~\cite{dumford2018backdooring,costales2020live} intuitively assume that the attacker somehow can access the model resided in the host file storage system or memory, e.g., through a toolkit, and tamper the weight after model deployment. It demonstrates the possibility of backdooring a model through perturbing the weights, though it is hugely computationally hungry to achieve a good attack successful rate, e.g., merely up to 50\%~\cite{dumford2018backdooring}. The main reason is an extensive iterative search over the weights that will be tampered for the sake of optimizing the ASR~\cite{dumford2018backdooring}. Costales {\it et al.}~\cite{costales2020live} takes the evasion defense as an objective when searching for parameters to be modified to bypass existing online inspection defense, specifically, evading the STRIP~\cite{gao2019strip} (STRIP is detailed in \autoref{sec:backdoorCountermeasure}).

{\it Bit Flip.} Rakin {\it et al.}~\cite{rakin2020tbt} recognize the advantage of flipping vulnerable bits of DNN weights to introduce backdoor. To determine which bits need to be flipped, the last-layer neurons with the most impact on the output for the targeted class are found using a gradient ranking approach. This eventually requires knowledge of both model architecture and parameters, which could be empowered by model extraction attack~\cite{jagielski2020high}. In this context, it is worth mentioning that recent microarchitectural side-channel attacks show the feasibility of gaining such information, including revealing architecture~\cite{yan2018cache} and stealing parameters~\cite{van2020cacheout} in the cloud with the existing shared covert-channel. The trigger is then generated using a minimization optimization technique. Then, the original weight matrix and the final optimized malicious weight matrix are compared---the weight here is quantified to 8-bit, providing the information on which bits need to be flipped. A small set of test samples is required to determine the malicious weight matrix. The backdoor is inserted through a row-hammer attack to flip the identified bits in main memory---but this step is not really implemented in~\cite{rakin2020tbt}. In one experiment, the authors~\cite{rakin2020tbt} can achieve backdoor effect with a small number of bit flips, e.g., 84 out of 88 million weight bits on ResNet18~\cite{he2016deep} on CIFAR10. It is acknowledged that, when backdooring a model trained over the large-scale ImageNet using this bit flip approach, the CDA and ASR drops are relatively noticeable---the attacking task becomes harder to the attacker. On the other hand, such post-deployment attacks via bit flipping---a kind of fault injection attack---are not restricted to {\it backdoor attacks}. It can be used to degrade the model overall performance significantly ~\cite{rakin2019bit,hong2019terminal,yao2020deephammer}, e.g., making all inputs accuracy to be low to random guess that is akin to denial of service. In this case, the bits can be {\it randomly flipped}. 

{\it TrojanNet.} Guo {\it et al.}~\cite{guo2020trojannet} propose TrojanNet that eventually switches the backdoored model from its public/main application task, e,g., traffic sign recognition, to a {\it different} secret application task, e.g., face recognition given the presence of the trigger. This is a different backdoor strategy, since the public task and secret task are for {\it totally differing applications}. For all other backdoor attacks, the attacker's task and the default benign task are {\it within the same application}. The public task and secret task of TrojanNet share no common features, and the secret task remains undetectable without the presence of a hidden trigger---here is a key~\cite{guo2020trojannet}. Generally speaking, the secret task and public task share the {\it same model parameter}. Switching from public task to secret task is realized by parameter permutation, which is determined and activated by a key/trigger. As stated by the authors, parameter permutation can be done at loading time or run-time in memory when a trigger is activated. Such a trigger can be embedded in malicious software. Training the public task and secret together is still akin to multi-task learning, though they do not share any feature. It is shown that multiple secret tasks can co-exist with the public task, while the utility of all secret tasks and the public task can be retained. Such a backdoor attack is tough to detect.


{\bf Notes.} \texttt{Post-deployment} backdoor attacks tend to be non-trivial to perform by the attacker because it requires first intruding the system to implant a malicious program~\cite{guo2020trojannet,dumford2018backdooring} or flipping the vulnerable bits in memory during run-time (as a type of fault injection) that needs professional expertise~\cite{rakin2020tbt}. In addition, \texttt{Post-deployment} backdoor attack usually requires white-box access to the model. The advantage of this attack is that it bypasses all offline inspections on the model or/and the data.


\subsection{Code Poisoning Attack}
Security risks of ML framework building upon public repositories or libraries have been initially recognized in~\cite{xiao2018security}. Regarding the backdoor attack, Bagdasaryan {\it et al.}~\cite{bagdasaryan2020blind} studied poisoning the code performing loss computation/function that could be exploited by an attacker to insert a backdoor into the model, which is recognized as a new type of backdoor attacks. 

Backdoor insertion is viewed as an instance of multi-task learning for conflicting objectives that are training the same model for high accuracy on the {\it main} and attacker-chosen {\it backdoor task} simultaneously. In this context, the attacker can take the i) main task, ii) backdoor task, and iii) even defense-evasion objectives into a single {\it loss function} to train the model. Bagdasaryan {\it et al.}~\cite{bagdasaryan2020blind} take advantage of a {\it multiple gradient descent algorithm} with the Franke-Wolfe optimizer to construct an optimal, self-balancing loss function, which achieves high accuracy on both the main and backdoor tasks. It is also noticed that auditing\footnote{The codebase of loss computation code could contain dozens of thousands of
lines, which are hard to understand even by experts.} and testing the loss-consumption code is hard in practice~\cite{bagdasaryan2020blind}. This backdoor is therefore blind to the attacker, termed as {\it blind backdoor attack}, since the attacker has no access to the later poisoned code during its execution, nor the training data on which the training operates, nor the resulting model, nor any other output of the training process (e.g., model accuracy). To be both data and model-independent considering that neither is known in advance, trigger inputs are synthesized ``on the fly''. Similarly the corresponding backdoor loss is also computed ``on the fly''. In this manner, the backdoor loss is always included in the loss function to optimize the model for both the main and backdoor tasks.

It is further demonstrated~\cite{bagdasaryan2020blind} that the backdoor task can be totally different from the primary task. For example, the main task of a model is to count the number of faces in a photo, while the backdoor task is covertly to recognize specific individuals. Similar backdoor attack effect---switching application---has been shown through TrojanNet~\cite{guo2020trojannet}. In addition, multiple backdoor tasks can be achieved using multiple synthesizers concurrently~\cite{bagdasaryan2020blind}. Moreover, evasion objective has been taken into the loss function to show evasion of various countermeasures~\cite{bagdasaryan2020blind}---this is an adaptive attack.

{\bf Notes.} We regard the \texttt{code poisoning} backdoor attack is with the widest victims as it tampers the low-level ML code repository. Such an attack even requires no knowledge of the model and no access to training data but achieves high ASR. 

\subsection{Discussion and summary}
Table~\ref{tab:ThreatModel} compares attacking capabilities and summarizes works under each attacking surface. Most of them attack classification task, especially, image classification. We can also observe that majority backdoor attacks are resulted from \texttt{outsourcing}, \texttt{pretrained}, \texttt{data collection}, \texttt{collaborative learning} attacking surfaces. Backdoor attack studies from \texttt{code poisoning} and \texttt{post-deployment} attack surfaces are relative less.

Unsurprisingly, \texttt{outsource} has the most robust control of the trigger and the highest ASR because the attacker has the maximum capability, including access to the model, training data, and the training process control. In contrast, it is worth noting that the defender/user is with the minimum capability when devising countermeasures under such \texttt{outsourcing} attacks. Because the defender is usually limited to ML expertise and computational resources, which are the reasons for outsourcing, the countermeasures should adequately consider the defenders' weakest capability. This consideration should also be always held when countering \texttt{pretrained} attacks because the user may usually be lack of computational resource.

Though rare studies under \texttt{code poisoning}, it appears that it has the widest attacking victims as it requires minimum attacking capability. The \texttt{pretrained} and \texttt{data collection} attack also have wide attacking victims. Except for \texttt{data collection}, the attacker is very unlikely to expose trigger inputs or poisoned data to defender until the trigger has to be presented at {\it attacking phase}. This explains the offline data inspection is inapplicable against attack surfaces except for \texttt{data collection}. It is very challenging for countering \texttt{collaborative learning} attacks because there is a trade-off between data privacy and security vulnerabilities. The defender, e.g., server, is even not allowed to access any testing/training data to assist the defense. In this case, the only countermeasure~\cite{chen2019deepinspect} requiring no testing data can be potentially considered. However, this countermeasure~\cite{chen2019deepinspect} first reverses engineers' training data, partially violating the purpose of adopting collaborative learning to protect data privacy. In addition, it is impossible to guarantee that none of the heterogeneous resided clients of collaborative learning is malicious as the number of clients could be high, e.g., hundreds of thousands. For the \texttt{pretrained} attack, it is worth to mention that some knowledge of the downstream student model tasks or/and training data is usually necessary to perform the attack. For both \texttt{pretrained} and \texttt{post-deployment} attacks, the trigger is usually delicately generated through some optimization algorithm, e.g., gradient-based approach, rather than arbitrarily chosen in order to achieve high ASR. Therefore, the attacker's flexibility of crafting arbitrary trigger is, more or less, deteriorated.

Moreover, we recommend that attack studies always explicitly identify their threat models to clarify the attacker's capabilities. 

\section{Backdoor Countermeasures}\label{sec:backdoorCountermeasure}

\begin{figure*}
	\centering
	\includegraphics[trim=0 0 0 0,clip,width=\linewidth]{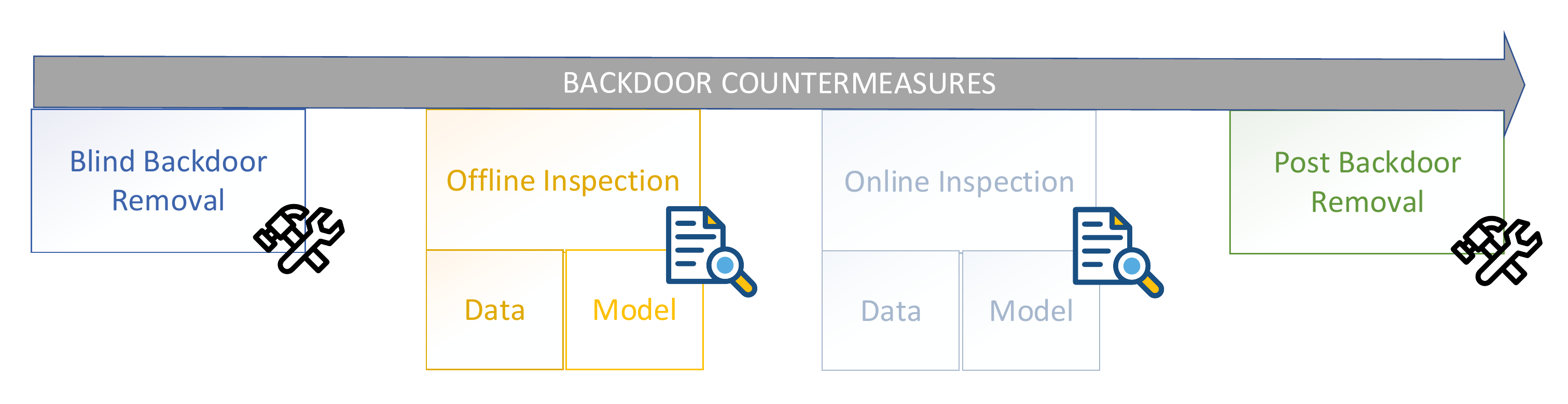}
	\caption{ Categories of backdoor countermeasures: blind backdoor removal without first check whether the model is backdoored or not; offline inspection on the data or model; online inspection on the (upcoming) data or model; post backdoor removal or restoration after backdoor trigger is identified.}
	\label{fig:Countermeasure}
\end{figure*}

We firstly categorize countermeasures into four classes: blind backdoor removal, offline inspection, online inspection, and post backdoor removal, as illustrated in Fig.~\ref{fig:Countermeasure}. These four classes are, somehow, align with possible sequential defense steps that could be considered. We performed a systematic analysis of countermeasures under these four classes. At the end of this section, we provide comparisons, summaries, and discussions.



\subsection{Blind Backdoor Removal}
The main feature of this defense categorization is that it does not differentiate the backdoored model from a clean model, or trigger input from clean input. The main purpose is to remove or suppress the backdoor effect while maintaining the CDA of clean inputs, especially considering that it is eventually a clean model. Notably, it is worth to mention that the blind backdoor removal can be performed offline or/and online.

{\it Fine-Pruning.} 
As an earlier defense, Liu {\it et al.} propose to remove potential backdoor by firstly pruning carefully chosen neurons of the DNN model that contribute least to the main classification task~\cite{liu2018fine}. More precisely, neurons are sorted by their activation on clean inputs (e.g., the hold-out validation samples) and pruned in the least-activated order. One general assumption is that the neurons activated by clean and trigger inputs are different or separable. After pruning, fine-tuning is used to restore model performance. Nonetheless, this method substantially degrades the model accuracy~\cite{wangneural}. It is also cumbersome to perform fine-pruning operations to any DNN model, as most of them are more likely to be benign. A similar approach presented in~\cite{liu2017neural} incurs high complexity and computation costs. 

{\it Februus.} Doan {\it et al.} propose Februus~\cite{februus} to eliminate the backdoor effect by online detecting and to sanitize the trigger input at run-time. It can serve as a filter deployed in front of any neural networks to cleanse the (trigger) inputs. The idea is to utilize a Visual Explanation tool to recognize the potential trigger region. Once the region is identified, Februus will surgically remove the malicious area and replace it with a neutrally gray color. This surgically removing step has already eliminated the trigger effect and potentially degrade the classification of the deep neural networks. One additional step, image restoration, was proposed using a GAN-based image inpainting method to restore the damaged areas to its original likeliness before being contaminated. This method is robust against multiple backdoor variants, especially the class-specific backdoor variant. However, as the method involves removing and restoring the images, it is sensitive to a large-size trigger covering a considerable portion of the images. 

{\it Suppression.} Sarkar {\it et al.}~\cite{sarkar2020backdoor} propose a backdoor suppression via fuzzing that entails building a wrapper around the trained model to neutralize the trigger effect. Given each input, many replicas are made: each replica is perturbed by adding some noise---noise level is empirically determined~\cite{sarkar2020backdoor}. All perturbed replicas are fed into the DNN model, regardless of backdoored or clean, to collect predictions. It is worth to mention such a strategy is previously exploited by~\cite{gao2019strip}. The final prediction of the input is based on the majority voting of predictions of its perturbed replicas. The evaluations on MNIST and CIFAR10 dataset demonstrate~\cite{sarkar2020backdoor} that around 90\% and 50\% of trigger images are reverted to their true predictions, respectively, which are less effective than other run-time inspection methods, e.g.,~\cite{gao2019strip}. On the other hand, it is acknowledged that this method compromises the CDAs for clean inputs relatively unacceptable~\cite{sarkar2020facehack}. Thus, it is less suitable as a wrapper around the trained model. In addition, the parameters such as the level of noise and the number of replicas that define this wrapper are empirically searched and determined, not generic. Moreover, this method appears to be reasonable only when the trigger is quite small\footnote{The authors evaluate triggers covering at most 5 pixels~\cite{sarkar2020facehack} for the MNIST dataset.}. Under such a condition only, fuzzing the inputs could suppress the trigger effect on hijacking the prediction.


{\it ConFoc.} The ConFoc~\cite{villarreal2020confoc} enforces the (clean or backdoored) model to focus on the content of the input image to remove backdoor effects while discarding the style information potentially contaminated by the trigger. It produces true labels---not the attacker targeted label---of the trigger input that is consistent with the input content, which retains the CDA with clean inputs as well. The rationale is that the image consists of both content and style information~\cite{gatys2016image}. The former refers to the shapes of the object or the semantic information, while the latter refers to the colors or texture information of images. It is hypothesized that focusing on content only is similar to human reasoning when making classifications. Therefore, the ConFoc~\cite{villarreal2020confoc} retrains the model to make classifications, mainly relying on the content information. One main limitation of the ConFoc is that it assumes the trigger does not overlap with the interested object---with content information---in the input image, which is inapplicable when setting triggers being overlapped with the object, e.g., those triggers in Fig.~\ref{fig:TriggerType}.

{\it RAB.} Weber {\it et al.}~\cite{weber2020rab} propose a provably robust training process for classifiers/models against backdoor attacks---a simpler concurrent work is in~\cite{wang2020certifying}. Generally, it~\cite{weber2020rab} exploits randomized smoothing to mitigate the trigger effects. It is worth to note that certified robustness is originally devised to against adversarial examples~\cite{lecuyer2019certified}. More precisely\footnote{We informally describe it to ease the general understanding.}, firstly, given the poisoned data, each sample is added certain noise. This creates one so-called smoothed training set. Such process is repeated to create a number of, e.g., 1000, replicated smoothed training set. Secondly, each smoothed training set is used to train a model, during training process, differential privacy can be further applied to improve the `smoothness' of the trained model. Therefore, a number of, e.g., 1000, models need to be trained. Thirdly, during inference time, an input $x$ is randomly perturbed according to certain noise bound to make replicas. All of $x$ purturbed replicas are fed into {\it each} trained model to gain the prediction label according to e.g., majority voting. Then the prediction labels from each model---in total, e.g., 1000 models---is further aggregated and the final label of $x$ is produced according a criteria, e.g., again majority vote. The purpose is that the final label will be the correct label that the clean model produces given the trigger perturbation is bounded. Such robustness can be proved. Though the provable robustness is desirable, the RAB~\cite{weber2020rab} has strict conditions such as the trigger perturbation bound must be small, which means that it can be easily bypassed in practice whenever the perturbation of the trigger exceeds the bound. In addition, there are many `smooth' models required to be trained, which greatly increases the computational overhead. Moreover, to guarantee the provable robustness, the RAB inexplicitly needs to have knowledge of the fraction of the poisoned data and the perturbation bound type, e.g., $l_2$-norm, which tend to be not practical. Notably, the RAB assumes access to poisoned data. As this countermeasure, to a large extent, is performed online that is to eliminate the trigger effect given an input $x$, so that it turns to be cumbersome when deployed for real-time applications as it could have relative unacceptable latency.


{\bf Notes.} Blind backdoor removal does not tell the backdoored model from a clean model.  As we can see, fine-pruning usually renders to CDA drop that could be unacceptable. Other works suppress or remove trigger effects if the model is infected. However, we note that suppression makes, somehow, the unrealistic assumption about the trigger size. As for ConFoc, it is only applicable to specific tasks where the image consists of both content and style information. Except for fine-pruning, all other works are inapplicable to other domains. They are all devised explicitly for the vision domain.

\subsection{Offline Inspection}
Both {\it backdoored model} and {\it trigger input} (or poisoned data) can be inspected offline. However, poisoned data inspection is only feasible under \texttt{data collection}. It is implausible for other attack surfaces where the attacker will not expose the poisoned data to the defender.  
\subsubsection{Data Inspection} Data inspection assumes that the poisoned data are available for the defenders since the attacker exploits data poisoning in order to insert backdoors.

{\it Spectral Signature.} Tran {\it et al.}~\cite{tran2018spectral} explore spectral signature, a technique based on (robust) statistic analysis, to identify and remove poisoned data samples from a potentially compromised training dataset. Firstly, a model is trained on a collected dataset that can contain poisoned data points. Secondly, for each particular output class/label, all the input instances for that label are fed into the model, and their {\it latent representations} are recorded. Thirdly, singular value decomposition is performed on the covariance matrix of the latent representations extracted, e.g., from the penultimate layer, which is used to compute an outlier score for each input. Fourthly, the input sample with a higher score than a ration is flagged as a trigger input, and removed from the training dataset on which a clean model will be trained again. This defense succeeds when the latent representations of clean inputs are sufficiently different from that of trigger inputs. Therefore, a crucial step is determining the means of obtaining proper latent representation to expose the trace of the trigger~\cite{ramakrishnan2020backdoors}. However, it points out that the outlier ration is fixed to be close to the ratio of corrupted samples in the target class. This requires some knowledge of the poison ratio and target class~\cite{chan2019poison,xiang2019benchmark}, which turns to be unknown in practice. 

{\it Gradient Clustering.} Chan {\it et al.}~\cite{chan2019poison} use cluster concept in a slightly different means that is not clustering {\it latent representations}. Chan {\it et al.} hypothesize and then show that a trigger image sample could lead to {\it a relatively large absolute value of gradient in the input layer} at the trigger position. Based on this, trigger samples can be separated from clean samples using a clustering algorithm. As the authors can detect the infected targeted and source classes, the infected model can be retrained by relabeling the trigger samples to their correct label to unlearn the backdoor effect, rather than training the model from scratch using sanitized data that has removed the corrupted data.  

{\it Activation Clustering.} Chen {\it et al.}~\cite{chen2018detecting} propose an activation clustering (AC) method. It is noticed that activations of the last hidden layer reflect high-level features used by the neural network to reach a model decision. Given the collected data and the model trained with the data, each sample is fed into the model, and the corresponding activation is collected. The activations of inputs belonging to the same label are separated and clustered by applying $k$-means clustering with $k=2$ after dimension reduction. Since the 2-means clustering will always separate the activations into two clusters, regardless of whether poisoned data is present or not, some metric to judge is used. For example, a high {\it silhouette score} means that the class has been infected or poisoned. Once the poisoned data are identified, they can be removed, and a clean model is trained. The other model restoration is relabeling the corrupted data to their correct source label to retrain the backdoored model to unlearn the backdoor effect, which could save time. 

{\it Deep k-NN.} In~\cite{peri2019deep}, Peri {\it et al.} design a deep k-NN method to detect {\it clean-label poisoned} samples, which can effectively against both feature collision and convex polytope clean-label attacks, see \autoref{sec:TM3Attack}. It demonstrates to detect over 99\% of poisoned examples in both clean-label attacks over the CIFAR-10 dataset. Consequently, those detected poisoned samples are removed without compromising model performance. Particularly, it has been shown more robust than $l_2$-norm, one-class SVM, and Random Point Eviction defenses against Feature Collision Attack, while comparable with $l_2$-norm defense but remove less clean images in Convex Polytope Attack.

{\it SCAn.} Unlike most countermeasures focus on class-agnostic triggers, Tang {\it et al.}~\cite{tang2019demon} propose a statistical contamination analyzer (SCAn), as a means of detecting both class-agnostic and class-specific triggers, see \autoref{sec:TriggerVariant}. It decomposes the latent representation of an image, in particular, the feature from the last third layer or the layer before logits, into two components: a class-specific identity, e.g., facial features distinguishing one person from others in face recognition task (that is a between-class component), and a variation component, e.g., size, orientation (within-class variation component). Then it is hypothesized that the variation component of an input image is independent of the label. This means the variation component distribution learned from one label/class is transferable to other labels. For example, smile as a variation component in face recognition is dependent on an individual's identity---the variation component is universal. Tang {\it et al.}~\cite{tang2019demon} decompose all images belonging to each class to obtain {\it finer-grained information}---in comparison with, e.g., the information gained through simple activation clustering~\cite{chen2018detecting}---on trigger impacts for classification. Using statistical analysis on the decomposed components, it is noted that the latent representation of images in the attacker target class (infected class) is a mixture of two groups from trigger inputs and clean inputs. Each is decomposed into distinct identity and universal variation components. Recall that the trigger images are from a different class/identity but (mis)labeled to the target label. So that the infected class can be identified. The SCAn has one additional main advantage that is also applicable to class-specific triggers. As a slight contradiction, it eventually assumes a small set of hold-out clean validation set, must contain no trigger, from in-house collection under this \texttt{data collection} surface. Although this assumption might be held in some cases. In addition, the SCAn is specific to image classification tasks where a class label is given to each object (face, flower, car, digits, traffic sign, etc.), and the variation applied to an object (e.g., lighting,
poses, expressions, etc.) {\it is of the same distribution across all labels}. Such a criterion may not always be satisfied. Moreover, it is acknowledged~\cite{tang2019demon} that SCAn is less effective to multiple target-trigger attack, see \autoref{sec:TriggerVariant}.

{\it Differential Privacy.} Du {\it et al.} propose an entirely different strategy that applies differential privacy when performing model training to facilitate the detection of outliers~\cite{du2019robust}. Applying differential privacy gains a trained model that is also named as a naturally smoothed model~\cite{weber2020rab}. This strategy can be effectively used to detect poisoned data because poisoned data can be viewed as outliers. Differential privacy is a means of protecting data privacy. It aims to hide specific input data from the output so that one can not tell whether the input data contain a particular record or not by looking at the result calculated from input data. When differential privacy is applied to the machine learning algorithm, the trained model becomes {\it insensitive to the removal or replacement of an arbitrary point in the training dataset}. More specifically, the contribution from rarely poisoned samples will be hidden by random noise added when training the model, so that the model underfits or suppresses the poisoned samples. This means the backdoor effect is already greatly suppressed by the model---ASR is quite low. In addition, the model will be less confident when predicting atypical poisoned examples. Therefore, by measuring the loss as a metric, one can distinguish poisoned inputs because it has a higher loss score. It is shown that this differential privacy based backdoor countermeasure~\cite{du2019robust} can effectively suppress backdoor effects as well as to detect poisoned samples, even the source-specific trigger inputs. As the validation is only through the toy MNIST dataset, it is worth further validating the efficacy of this method with more complex datasets trained by representative model architectures.



{\bf Notes:} We can see that most of these countermeasures are based on a clustering algorithm applied on, e.g., latent representation, gradient, and activation. We note that these countermeasure may not be applicable for the special image-scaling based backdoor attack under \texttt{data collection} attack surface (see~\autoref{sec:TM3Attack}). To defend such an attack, we need to carefully choose the downsampling filter parameters during resizing to remove the image-scaling attack effect~\cite{quiring2020adversarial}. For the other special clean-label poisoning attack that is the targeted class data poisoning by only stamping triggers on the data belonging to the targeted class~\cite{barni2019new}, it is unclear whether the above data inspection methods are effective or not. Because they assume the source class is a different target, violating the attack setting of~\cite{barni2019new} that source class stamped with the trigger is the same as the target class. In addition, most studies do not explicitly consider reporting results of, e.g., false positive, when the dataset under examination is eventually benign.

\subsubsection{Model Inspection}\label{sec:offlineModelInspection}

Unlike data inspection that assumes access to poisoned data. All countermeasures in this part use a more realistic assumption that the poisoned data is not available. Therefore, they are suitable for countering backdoor attacks resulted from various attacking surfaces---not only limited to \texttt{data collection}.

{\it Trigger Reverse Engineer.} In Oakland 2019, Wang {\it et al.}~\cite{wangneural} propose the NeuralCleanse to detect whether a DNN model has been backdoored or not prior to deployment, where the performance is further improved in TABOR~\cite{guo2019tabor} by utilizing various regularization when solving optimizations. The other improvement with a similar concept of NeuralCleanse is studied in~\cite{xiang2019revealing}, although the specific methods can be varied. NeuralCleanse is based on the intuition that, given a backdoored model, it requires much smaller modifications to {\it all input samples} to misclassify them into the attacker targeted (infected) label than any other uninfected labels. In other words, a {\it shortest perturbation path} exists to {\it any input} to the target label. Therefore, NeuralCleanse iterates through all labels of the model and determines if any label requires a substantially smaller modification to achieve misclassifications.
One advantage of this method is that the trigger can be reverse-engineered and identified during the backdoored model detection process. Once the trigger is identified, backdoor removal (detailed in Section~\ref{sec:RemovalBackdoor}) can be performed via retraining to unlearn the trigger effect. 

Despite its novelty, NeuralCleanse is still with some limitations. Firstly, it could incur high computation costs {\it proportional to the number of labels}. The computation cost of the detection process can take several days for specific DNN models, even when the optimization is considered. This is especially the case when the number of classes of the task is large. Secondly, like most backdoor countermeasures, the method is reported to be less effective with the increased trigger size. Thirdly, it requires training reference models to determine a (global) threshold to distinguish clean models from backdoored models, which inadvertently appear to be inapplicable under \texttt{outsourcing}. Because this violates the motivation of \texttt{outsourcing} where the user has limited computational resource or/and ML expertise.

It is observed that the reversed trigger is not always consistent with the original trigger\cite{qiao2019defending,veldanda2020nnoculation,harikumar2020scalable}, e.g., in different runs~\cite{qiao2019defending}. Consequently, when the reversed trigger that is inconsistent with the original trigger is utilized for backdoor removal, the removal efficacy will be significantly decreased. That is, the input with the original trigger still has a considerable attack success rate to the retrained model. Accordingly, Qiao {\it et al.}~\cite{qiao2019defending} generatively model the valid trigger distribution, rather than a single reversed trigger, via the max-entropy staircase approximator (MESA) and then use many valid triggers to perform backdoor removal to increase the removal efficacy. However, this work relies on several {\it strong assumptions}, such as knowledge of the model being backdoored and the shape and size of the trigger, which are not available in real-world to defenders. In addition, validations in~\cite{qiao2019defending} are limited by a small trigger, a $3\times3$-pixel square, where the concerned computational overhead of this work is, however, not reported.

Harikumar {\it et al.}~\cite{harikumar2020scalable} propose Scalable Trojan Scanner (STS) in order to eliminate one limitation of the NeuralCleanse: the computational overhead is linearly increased with the number of classes in the dataset. The STS reverse engineers the trigger in a means of identifying a trigger (perturbation) that results in prediction vectors, e.g., softmax, of {\it all images} to be similar. In this context, the trigger searching (reverse engineer) is {\it independent} on the number of classes (this is where the {\it Scalable} comes from). In other words, it does not care about the target class when reverse engineering the trigger. The STS saves time in comparison with NeuralCleanse when searching the minimum perturbation---the identified trigger---through optimization algorithm. However, in~\cite{harikumar2020scalable}, it only tests triggers with {\it small size and fixed shape} (square). It appears that the STS is still ineffective to large trigger size. 

{\it NeuronInspect.} Under similar concept with SentiNet~\cite{chou2018sentinet} and Februus~\cite{februus} exploiting {\it output explanation}, Huang {\it et al.}~\cite{huang2019neuroninspect} propose NeuronInspect to combine output explanation with outlier detection considering that the explanation heatmap of attacker target class---treated as outlier---differs from non-target class. In other words, the target class's heatmap is hypothesized to be i) compact, ii) smooth, and iii) remains persistent even across different input images. The persistence is an exhibition of the trigger input-agnostic characteristic. In comparison with NeuralCleanse, NeuronInspect has advantages, including reduced computational overhead and robustness against a multiple-trigger-to-same-label backdoor variant (V1 in \autoref{sec:TriggerVariant}). However, unlike NeuralCleanse~\cite{wangneural}, TABOR~\cite{guo2019tabor} and STS~\cite{harikumar2020scalable}, NeuronInspect does not reverse engineer and discovers the trigger.

{\it DeepInspect.} Chen {\it et al.} propose DeepInspect~\cite{chen2019deepinspect} to detect backdoor attack with {\it no requirement of accessing training data}, because it first reverse engineers training data. The key idea of DeepInspect is to use a conditional generative model to learn the probabilistic distribution of potential triggers. This generative model will be used to generate reversed triggers whose perturbation level will be statistically evaluated to build the backdoor anomaly detection. DeepInspect is faster than NeuralCleanse~\cite{wangneural} to reverse triggers in complex datasets. However, due to the weak assumption of the defender's capability that has no access to training data, the detection performance of DeepInspect appears to be worse than NeuralCleanse.

{\it AEGIS.} Soremekun {\it et al.}~\cite{soremekun2020exposing}, for the first time, investigate the backdoor attacks on {\it adversarial robust model that is elaborately trained defending against adversarial attacks}~\cite{santurkar2019image}. It is shown that the adversarial robust model, robust model for short in the latter description, is vulnerable to backdoor attacks, though the robust model is robust against adversarial example attacks. Correspondingly, they propose AEGIS, a method exploiting latent feature clustering to detect {\it backdoored robust model} accurately. One exploited the unique property of a robust model is that it can be adopted for image generation/synthesizing~\cite{santurkar2019image}. Therefore, for each class, the robust classifier {\it itself} is used to generate synthesized images. Then latent features of both synthesized images and held-out validation images are extracted, followed by feature dimension reduction. After that, a clustering algorithm, namely the shift algorithm~\cite{fukunaga1975estimation}, is used to automate the detection process. If there are no more than two clusters---one is the distribution of the validation samples, one is the distribution of the synthesized images uninfected by triggers, then this robust model is deemed as clean, otherwise backdoored because now there are other distributions formed by synthesized images infected by triggers. Clustering the input samples with their latent representations is not new, which has been used for trigger data inspection~\cite{chen2018detecting}. The difference is that previous work~\cite{chen2018detecting} requires to access poisoned trigger data, while AEGIS does not. The limitation of AEGIS is only applicable for robust models, not standard models. It is also acknowledged that defeating specific triggers, e.g., blending trigger as in Fig.~\ref{fig:TriggerType} (a), induced backdoor appears to be a challenge for AEGIS.

{\it Meta Classifier.} Unlike the above works dedicated to the vision domain, Xu {\it et al.}~\cite{xu2019detecting} consider a generic backdoor detection applicable to diverse domains such as audio and text. It trains many (e.g., 2048) clean and (e.g., 2048) backdoored shadow models by the defender to act as training samples of a meta-classifier---the other neural network---to predict whether a new model is clean or not. Note that the shadow model is sometimes called a surrogate model. To approximate the general distribution of backdoored models, the shadow backdoored models are trained over diverse trigger patterns and backdoor goals. To generate feature input to the meta-classifier, many query inputs are made to each shadow model, and confidence scores from the shadow model are concatenated, acting as feature representation of the shadow model. In contrast, the output of the meta-classifier is binary. The number of the query inputs is set as 10~\cite{xu2019detecting}. The query inputs can further be optimized along with the training of the meta-classier, rather than randomly selected to gain accurate feature representation of the shadow model~\cite{xu2019detecting}. This method has the advantage of its generalization across different domains. It is worth to mention that adaptive attacks are considered explicitly by the authors to evaluate the defense robustness. However, this method requires both ML expertise and expensive computational resources, which might be less practical.

{\it Universal Litmus Pattern.} Colouri {\it et al.}~\cite{kolouri2020universal} apply similar concept as above~\cite{xu2019detecting} to train a meta-classifier. Like~\cite{xu2019detecting}, a small set of specifically chosen image inputs, here namely, universal litmus patterns~\cite{kolouri2020universal}, query the clean and backdoored shadow models, then the logits (from the penultimate layer) are concatenated as feature input to the meta-classier, whether the model is clean or backdoored is the output of the meta-classifier. The universal litmus patterns are crafted with the assistance of the technique in~\cite{wangneural}. When a new model is coming, these universal litmus patterns are fed into it and collect the new model's logits as a feature to the meta-classifier to tell whether this new model is backdoored. Both~\cite{xu2019detecting} and \cite{kolouri2020universal} require train enormous clean and backdoored {\it models} as prerequisites, the cost is high. This would worsen because the meta-classifier is dataset dependent. For example, if the meta-classifier is for CIFAR10, a new model trained over MNIST cannot be distinguished by this meta-classifier. Therefore, such computational expensive shadow model training has to be performed when dataset changes. In addition, in comparison with~\cite{kolouri2020universal} that is generalize to diverse domains, the universal litmus pattern is specific to image inputs. Moreover, the efficacy of universal litmus pattern~\cite{xu2019detecting} is achieved with a strong assumption: the trigger size is inexplicitly assumed to be known. 

{\bf Notes.} As the defender is with less capability, in particular, no access to poisoned samples, it is unsurprising that offline model inspection usually requires high computational overhead and ML expertise. In this context, defenders under \texttt{outsourcing} and \texttt{pretrained} may not be able to adopt these countermeasures. Because they are usually lack of computational resources or/and ML expertise. For example, the countermeasures relying on meta classifier need train many shadow models that are computational hungry. In addition, their efficacy~\cite{xu2019detecting,kolouri2020universal} is potentially limited by the variety of the triggers used when training the meta classifier~\cite{harikumar2020scalable}. Most of the countermeasures cannot deal with large-sized triggers, especially for countermeasures aiming to reverse engineer the trigger. 


\subsection{Online Inspection}
Online inspection can also be applied to monitor the behavior of either the model or input data during run-time.

\subsubsection{Data Inspection} Data inspection checks the inputs to decide whether a trigger is contained or not through anomaly detection, thus, denying the adversarial input and then throwing alert if set.  

{\it SentiNet.} In 2018, Chou {\it et al.}~\cite{chou2018sentinet} exploit both the model interpretability/explanation and object detection techniques, referred to as SentiNet, to firstly discover contiguous regions of an input image important for determining the classification result. This region is assumed to have a high chance of possessing a trigger when it {\it strongly affects} the classification. Once this region is determined, it is carved out and patched on to other held-out images with ground-truth labels. If both the misclassification rate---probability of the predicted label is not the ground-truth label of the held-out image---and confidence of these patched images are high enough, this carved patch is regarded as an adversarial patch that contains a backdoor trigger. Therefore, the incoming input is a backdoor input.

{\it NEO.} Udeshi {\it et al.}~\cite{udeshi2019model} proposed NEO that is devised to search and then mitigate the trigger by the {\it dominant color} in the image frame when it appears in the scene at run-time. The method is claimed to be fast and completely black-box to isolate and reconstruct the trigger for defenders to verify on a skeptical model. However, by occluding the trigger with the dominant color, the method is not robust against large-sized triggers, and the classification accuracy after the NEO method is also degraded since the main feature is sometimes occluded. In addition, the validation is only performed via a trigger that is with a square shape, which is not generalized.


{\it STRIP.} In 2019 ACSAC, Gao {\it et al.}~\cite{gao2019strip} propose STRIP. They turn the input-agnostic strength of the trigger into a weakness exploited to detect the poisoned input. Suppose a model has been backdoored. On the one hand, for a clean input, the prediction $z$ should be quite different from the ground-truth given a strong and intentional perturbation is applied to it. Therefore, replicas of a clean input with varying perturbation exhibit strong randomness---quantified via the entropy metric. On the other hand, for trigger inputs, the prediction $z$ should usually be constant to the attacker's target $z_a$ even under perturbations because of the strong hijacking effect of the trigger. Therefore, replicas of the trigger input with varying perturbation exhibit weak randomness. Given a preset entropy threshold---can be determined solely using clean input, the trigger input (high entropy) can be easily distinguished from the clean input (low entropy), consequently, throwing alert for further investigations. The STRIP is simple to use, a non-ML method, yet efficient and generic, validated not only by the vision domain but also by text and audio~\cite{gao2019design}. It is robust against arbitrary triggers, e.g., any size---this is always a limitation for most countermeasures~\cite{chou2018sentinet,wangneural,guo2019tabor,qiao2019defending}. The STRIP concept is later adopted in~\cite{chen2020mitigating} to identify triggers in text for countering backdoor attacks on LSTM-based models, but under the \texttt{data collection} attacking surface. One limitation of STRIP is that it is mainly designed for class-agnostic triggers that would not be efficient for class-specific triggers.

{\it Epistemic Classifier.} Yang {\it et al.}~\cite{yang2020countermeasure} propose an epistemic classifier to leverage whether the prediction of input is reliable or not to identify inputs containing triggers---adversarial input is unreliable. The concept is similar to~\cite{papernot2018deep}. Generally, an input is compared to its neighboring training points---determined online---according to the specific distance metric that separates them in the latent representations across multiple layers. The labels of these searched neighboring training points in each hidden layer are used to check whether the intermediate computations performed by each layer conform to the final model’s prediction. If they are inconsistent, input prediction is deemed unreliable, which is a trigger input, otherwise, clean input. It is generally based on the hypothesis that input with trigger might start close to a clean training instance---from the source class---in the input layer, but its trajectory over the neural network will slowly or abruptly close to the attacker's chosen targeted class. This concept is similar to NIC (detailed soon in the below)~\cite{ma2019nic}. Such abnormal deviation can be reflected by observing the consistency of the labels of input's neighboring training points across multiple layers. Even though this work~\cite{yang2020countermeasure} is insensitive to trigger shapes, sizes, it results in high computational overhead when the model is with large number of convolutional layers or/and the dataset is large, which is problematic especially considering that it acts as an online backdoor detection approach.

\subsubsection{Model Inspection}
Model inspection~\cite{liu2019abs,ma2019nic} relies on an anomaly technique to distinguish a model's abnormal behaviors resulted from the backdoor.

{\it ABS.} In 2019 CCS, Liu {\it et al.} propose Artificial Brain Stimulation (ABS)~\cite{liu2019abs} by scanning a DNN model to determine whether it is backdoored. Inspiring from the Electrical Brain Stimulation (EBS) technique used to analyze the human brain neurons, Liu {\it et al.} use ABS to inspect individual neuron activation difference for anomaly detection of backdoor, this can potentially defeat backdoor attacks on sequential models~\cite{yang2019design} beyond classification tasks. ABS's advantages are that i) it is trigger size-independent, and ii) requires only one clean training input per label to detect the backdoor. iii) It can also detect backdoor attacks on feature space rather than pixel space. Nevertheless, the method appears to be only effective under certain {\it critical} assumptions, e.g., the target label output activation needs to be activated by {\it only one} neuron instead of from interaction of a group of neurons. Thus, it can be easily bypassed by using a spread trigger like the one in Fig~\ref{fig:TriggerType} (b) (see Section 5.2 in~\cite{veldanda2020nnoculation}). In addition, the scope is also limited to the attack of one single trigger per label. If multiple triggers were aimed to attack the same label, it would be out of ABS's reach. 

{\it NIC.} In 2019 NDSS, Ma {\it et al.} propose NIC~\cite{ma2019nic} by checking the provenance channel and activation value distribution channel. They extract DNN invariants and use them to perform run-time adversarial sample detection, including trigger input detection. This method can be generally viewed to check the activation distribution and flow across DNN layers---inspired by the control flow used in programming---to determine whether the flow is violated due to the adversarial samples. However, NIC requires extensive offline training to gain different classifiers across layers to check the activation distribution and flow and can be easily evaded by adaptive attacks~\cite{ma2019nic}.

\begin{table*}
	\centering 
	\caption{Backdoor countermeasures summary and comparison.}
			\resizebox{1.00\textwidth}{!}{
	\begin{tabular}{ c | c | c | | c |c | c | c | c | c | c | c | c | c | c } %
		\toprule 
		\toprule 
				
		\begin{tabular}{@{}c@{}} Categorization \end{tabular} & \begin{tabular}{@{}c@{}} Work \end{tabular} &  \begin{tabular}{@{}c@{}} Model \\ Access \end{tabular}  & \begin{tabular}{@{}c@{}} Domain \\ Generalization$^1$ \end{tabular} & \begin{tabular}{@{}c@{}} Poisoned \\ Data Access \end{tabular} & \begin{tabular}{@{}c@{}} Validation \\ Data Access \end{tabular} & \begin{tabular}{@{}c@{}} Computational \\ Resource \end{tabular} & \begin{tabular}{@{}c@{}} ML \\ Expertise \end{tabular} & \begin{tabular}{@{}c@{}} Global \\ Reference \end{tabular} &  \begin{tabular}{@{}c@{}} V1 \end{tabular} & \begin{tabular}{@{}c@{}} V2 \end{tabular} & \begin{tabular}{@{}c@{}} V3 \end{tabular} & \begin{tabular}{@{}c@{}} V4 \end{tabular} & \begin{tabular}{@{}c@{}} V5 \end{tabular} \\ 
		\midrule
		
	\multirow{4}{*}{\begin{tabular}{@{}c@{}}  ~\\~\\~\\ Blind \\ Backdoor \\ Removal \end{tabular}} & 	\begin{tabular}{@{}c@{}} Fine-Pruning \\ 2018~\textcolor{blue}{\cite{liu2018fine}}$^2$ \end{tabular}  &  White-Box & \LEFTcircle &\Circle &  \CIRCLE & Medium & High & \LEFTcircle & \CIRCLE & \CIRCLE & \CIRCLE & \CIRCLE & \CIRCLE \\ \cline{2-14}
	
	& \begin{tabular}{@{}c@{}} Februus \\ 2019~\cite{februus} \end{tabular} &  White-Box & \Circle &\Circle &  \CIRCLE & Medium & High & \Circle & \CIRCLE & \CIRCLE & \CIRCLE & \LEFTcircle & \LEFTcircle \\ \cline{2-14}
	
	& \begin{tabular}{@{}c@{}} Suppression \\ 2020~~\textcolor{blue}{\cite{sarkar2020facehack}} $^3$ \end{tabular} &  Black-Box & \Circle &\Circle & \CIRCLE & Low & Medium & \Circle & \CIRCLE & \CIRCLE & \LEFTcircle & \CIRCLE & ---$^4$ \\ \cline{2-14}
	
	& \begin{tabular}{@{}c@{}} Certified \\ 2020~\textcolor{blue}{\cite{wang2020certifying}} \end{tabular} &  Black-Box & \Circle &\Circle & \CIRCLE & Low & Medium & \Circle & \CIRCLE & \CIRCLE  & \Circle & \CIRCLE & --- \\ \cline{2-14}

	& \begin{tabular}{@{}c@{}} RAB \\ (Certified) \\ 2020~\textcolor{blue}{\cite{weber2020rab}} \end{tabular} &  White-Box & \CIRCLE &\CIRCLE & \CIRCLE & High & High & \Circle & \CIRCLE & \CIRCLE  & \Circle & \LEFTcircle & \begin{tabular}{@{}c@{}} \CIRCLE \end{tabular} \\ \cline{2-14}  
	
	& \begin{tabular}{@{}c@{}} ConFoc \\ 2020~\cite{villarreal2020confoc} \end{tabular}  &  White-Box & \Circle &\Circle &  \CIRCLE & Medium & High & \Circle & \CIRCLE & \CIRCLE  & \CIRCLE & \LEFTcircle & \LEFTcircle \\ \hline

	\multirow{5}{*}{\begin{tabular}{@{}c@{}}  ~\\~\\~\\~\\ Offline \\ Data \\ Inspection \end{tabular}} & 	\begin{tabular}{@{}c@{}} Spectral \\ Signature \\ 2018~\textcolor{blue}{\cite{tran2018spectral}} \end{tabular}  &  White-Box & \CIRCLE &\CIRCLE &  \CIRCLE & Medium & Medium & \CIRCLE & \CIRCLE & \CIRCLE & \CIRCLE & \CIRCLE & \CIRCLE \\ \cline{2-14}
	
	& \begin{tabular}{@{}c@{}} Activation \\ Clustering \\ 2018~\cite{chen2018detecting} \end{tabular} &  White-Box & \CIRCLE &\CIRCLE &  \CIRCLE & Medium & Medium & \CIRCLE & \CIRCLE & \CIRCLE & \CIRCLE & \CIRCLE & \begin{tabular}{@{}c@{}} \Circle \\ \cite{tang2019demon}  \end{tabular}  \\ \cline{2-14}

   	& \begin{tabular}{@{}c@{}} Gradient \\ Clustering\\ 2019~\cite{chan2019poison} \end{tabular} &  White-Box & \begin{tabular}{@{}c@{}} \LEFTcircle \\ \cite{chen2020mitigating}\end{tabular} & \CIRCLE & \CIRCLE & Medium & Medium & \Circle & \CIRCLE & \CIRCLE & \CIRCLE & \CIRCLE & \CIRCLE \\ \cline{2-14}
   	
   	& \begin{tabular}{@{}c@{}} Differential \\ Privacy \\ 2019~\cite{du2019robust} \end{tabular} &  White-Box & \CIRCLE & \CIRCLE & \CIRCLE & Medium & Medium & \Circle & \CIRCLE & \CIRCLE & \CIRCLE & \CIRCLE & \CIRCLE \\ \cline{2-14}  
  
	& \begin{tabular}{@{}c@{}} SCAn \\ 2019~\cite{tang2019demon} \end{tabular} &  White-Box & \LEFTcircle &\CIRCLE &  \CIRCLE & Medium & Medium & \CIRCLE & \CIRCLE & \LEFTcircle & \CIRCLE & \CIRCLE & \CIRCLE \\ \hline
		
	\multirow{6}{*}{\begin{tabular}{@{}c@{}}  ~~ \\~~ \\~~\\~~ \\~\\~\\~\\ Offline \\ Model \\ Inspection \end{tabular}} &
	\begin{tabular}{@{}c@{}} Neural  \\Cleanse \\ 2019 ~\textcolor{blue}{\cite{wangneural}} \end{tabular} &  Black-Box & \Circle & \Circle & \CIRCLE & High & High & \CIRCLE & \Circle & \LEFTcircle & \Circle & \CIRCLE & \begin{tabular}{@{}c@{}} \Circle \\ \cite{tang2019demon}  \end{tabular} \\ \cline{2-14}
	
	& \begin{tabular}{@{}c@{}} TABOR \\ 2019~\cite{guo2019tabor} $^5$ \end{tabular} &  Black-Box & \Circle & \Circle & \CIRCLE & High & High & \CIRCLE & \Circle & \LEFTcircle & \Circle & \CIRCLE & \Circle \\ \cline{2-14}
	
	& \begin{tabular}{@{}c@{}} Xiang {\it et al.} \\ 2019~\cite{xiang2019revealing} $^5$ \end{tabular} &  Black-Box & \Circle & \Circle & \CIRCLE & High & High & \CIRCLE & \Circle & \CIRCLE & \Circle & \CIRCLE & \CIRCLE \\ \cline{2-14}

	& \begin{tabular}{@{}c@{}} STS \\ 2020~\cite{harikumar2020scalable} \end{tabular} &  Black-Box & \Circle & \Circle & \CIRCLE & High & High & \Circle & \Circle & \Circle & \Circle & \CIRCLE & \Circle \\ \cline{2-14}	
	
	&	\begin{tabular}{@{}c@{}} DeepInspect\\2019~~\textcolor{blue}{\cite{chen2019deepinspect}} \end{tabular} &  Black-Box & \Circle & \Circle & \Circle & High & High & \Circle & \Circle & \LEFTcircle & \Circle & \CIRCLE & \Circle \\ \cline{2-14}

	& \begin{tabular}{@{}c@{}} NeuronInspect \\ 2019 ~\cite{huang2019neuroninspect} \end{tabular} &  White-Box & \Circle & \Circle &  \CIRCLE & High & High & \CIRCLE & \Circle & \CIRCLE & \Circle & \CIRCLE & \Circle \\ \cline{2-14}
	
	&	\begin{tabular}{@{}c@{}} AEGIS\\2020~\cite{soremekun2020exposing} \end{tabular} &  White-Box & \Circle & \Circle & \CIRCLE & Medium & High & \Circle & \CIRCLE & \CIRCLE & \LEFTcircle & \CIRCLE & --- \\ \cline{2-14}
	
    & \begin{tabular}{@{}c@{}} Meta \\ Classifier \\ 2019~\cite{xu2019detecting} \end{tabular} &  Black-Box & \CIRCLE & \Circle &  \CIRCLE & \begin{tabular}{@{}c@{}} Very \\ High \end{tabular} & High & \CIRCLE & \CIRCLE & \CIRCLE & \CIRCLE & \CIRCLE & \LEFTcircle \\ \cline{2-14}
    
    & \begin{tabular}{@{}c@{}} Universal \\ Litmus \\ Pattern\\2020~\cite{kolouri2020universal} \end{tabular} &  White-Box & \Circle & \Circle &  \CIRCLE & \begin{tabular}{@{}c@{}} Very \\ High \end{tabular} & High & \CIRCLE & \Circle & \LEFTcircle & \Circle & \CIRCLE & \Circle \\ \hline
	
	\multirow{3}{*}{\begin{tabular}{@{}c@{}} ~\\~\\~\\ Online \\ Input \\ Inspection \end{tabular}} & \begin{tabular}{@{}c@{}} SentiNet \\ 2018 \cite{chou2018sentinet} \end{tabular}  &  White-Box & \Circle & \Circle &  \CIRCLE & Low & Medium & \Circle & \CIRCLE & \CIRCLE & \LEFTcircle & \CIRCLE & \begin{tabular}{@{}c@{}} \Circle \\ \cite{tang2019demon}  \end{tabular} \\ \cline{2-14} 

	& \begin{tabular}{@{}c@{}} NEO \\ 2019 ~\cite{udeshi2019model} \end{tabular} &  Black-Box & \Circle & \Circle &  \CIRCLE & Medium & Medium & \Circle & \CIRCLE & \CIRCLE & \Circle & \CIRCLE & \begin{tabular}{@{}c@{}} --- \\  \end{tabular}\\ \cline{2-14}
	
	& \begin{tabular}{@{}c@{}} STRIP \\ 2019 ~\textcolor{blue}{\cite{gao2019strip}} \end{tabular} &  Black-Box & \CIRCLE & \Circle &  \CIRCLE & Low & NO & \Circle & \CIRCLE & \CIRCLE & \CIRCLE & \CIRCLE & \begin{tabular}{@{}c@{}} \Circle \\ \cite{tang2019demon}  \end{tabular}\\ \cline{2-14}
	
	& \begin{tabular}{@{}c@{}} Epistemic \\  Classifier \\ 2020~~\textcolor{blue}{\cite{yang2020countermeasure}} $^6$ \end{tabular} &  White-Box & \LEFTcircle & \Circle &  \CIRCLE & High & Medium & \Circle & \CIRCLE & \CIRCLE & \CIRCLE & \CIRCLE & \CIRCLE \\ \cline{2-14}
	
	& \begin{tabular}{@{}c@{}} NNoculation\\2020~\cite{veldanda2020nnoculation} \end{tabular} &  White-Box & \Circle & \Circle &  \CIRCLE & High & High & \LEFTcircle & \CIRCLE & \CIRCLE & \CIRCLE & \CIRCLE & \LEFTcircle \\ \hline

	\multirow{2}{*}{\begin{tabular}{@{}c@{}}   Online \\ Model \\ Inspection \end{tabular}} & \begin{tabular}{@{}c@{}} ABS\\2019~~\textcolor{blue}{\cite{liu2019abs}} \end{tabular} &  White-Box & \LEFTcircle & \Circle & \LEFTcircle & High & High & \CIRCLE & \Circle & \CIRCLE & \Circle $^7$ & \CIRCLE & \begin{tabular}{@{}c@{}} \Circle \\ \cite{sarkar2020facehack}  \end{tabular} \\ \cline{2-14}

	& \begin{tabular}{@{}c@{}} NIC \\  2019~\textcolor{blue}{\cite{ma2019nic}} \end{tabular} &  White-Box & \LEFTcircle & \Circle &  \CIRCLE & High & High & \Circle & \CIRCLE & \CIRCLE & \CIRCLE & \CIRCLE & \CIRCLE \\ \hline
	
	\bottomrule
	\end{tabular}}
	  \begin{tablenotes}
      \footnotesize
      \item \CIRCLE: Applicable or Necessary. \Circle: Inapplicable or Unnecessary. \LEFTcircle: Partially Applicable or Necessary.
      \item V1: Multiple Triggers to Same Label. V2: Multiple Triggers to Multiple Labels. V3: Trigger Size, Shape and Position. V4: Trigger Transparency. V5: Class-specific Trigger. See detailed descriptions about V1-V5 in Section~\ref{sec:TriggerVariant}.
      \item $^1$ Most defenses are specifically designed for the vision domain, which may not be applicable to other domains such as audio and text.
      \item $^2$ Fine-pruning is performed without firstly checking whether the model is backdoored or not, which could greatly deteriorate CDA both clean and backdoored models.
      \item $^3$ This work~\cite{sarkar2020facehack} is only suitable for {\it very} small triggers.
      \item $^4$ --- means such information is neither inexplicitly (hard to infer according to the work description) nor explicitly unavailable.
      \item $^5$ TABOR is an improvement over NeuralCleanse for improving the reverse engineered trigger accuracy. While~\cite{xiang2019revealing} share a concept similar to NeuralClease although the specific methods are different.
      \item $^6$ This work~\cite{yang2020countermeasure} becomes impractical for models with large convolutional layers or/and a large dataset, especially as an online model inspection approach.
      \item $^7$ It is shown that ABS is sensitive to the shape of triggers, e.g., spread or distributed triggers~\cite{veldanda2020nnoculation}.

    \end{tablenotes}
	\label{tab:countermeasure} 
\end{table*}

{\bf Notes:} It is worth to mention that online inspection indeed requires some preparations performed offline, for example, determining a threshold to distinguish the trigger behavior from clean inputs. However, the threshold is determined solely by clean inputs. One advantage of online inspection is that some countermeasures~\cite{ma2019nic,gao2019strip} are insensitive to trigger size. In addition, online inspection countermeasures~\cite{ma2019nic,gao2019strip,liu2019abs,yang2020countermeasure} has, to some extent, good generalization to diverse domains. One limitation is that the online inspection often results in latency. Therefore, online detection should be fast to be suitable for real-time applications such as self-driving.  

\subsection{Post Backdoor Removal}\label{sec:RemovalBackdoor}
Once backdoor, either via model inspection or data inspection, is detected, backdoor removal can be taken into consideration. One way is to remove the corrupted inputs and train the model again~\cite{tran2018spectral,chen2018detecting}, which appears to be only practical under \texttt{data collection} as the user is not allowed to access trigger inputs under other attack surfaces. 

The other way is to remove the backdoor behaviors from a model by retraining or fine-tuning the backdoored model using corrupted data comprising the trigger but labeled correctly, which relearns the corrected decision boundaries~\cite{veldanda2020nnoculation,chan2019poison}. In this context, the crux is to reverse engineer the original trigger precisely and identify the infected target label. NeuralCleanse~\cite{wangneural} is a representative of this kind. However, the reverse-engineered trigger often fails to be the original trigger~\cite{qiao2019defending,veldanda2020nnoculation}, which renders inefficiency of lowering the ASR of the corrected backdoored model. The ASR of the corrected model could still be high, e.g., around 50\% in the worst case (on CIFAR10 and the original trigger with the size of $3\times 3$ and simply black-white pattern), given the poisoned input~\cite{qiao2019defending}.

Qiao {\it et al.}~\cite{qiao2019defending} recognize that the reversed trigger follows a distribution instead of a single point so that a number of valid triggers under the trigger distribution space are used to combine with clean data to retrain the backdoored model. Here, these valid triggers are reversed from the original trigger but are visually different from the original trigger. However, they are still able to achieve the original trigger effect in some instances. The corrected model exhibits a better backdoor removal efficacy---ASR is significantly reduced to 9.1\% in the worst case (on CIFAR10 and the original trigger with the size of $3\times 3$ and simply black-white pattern). However, the critical limitation of this work is that it assumes knowledge of i) model being backdoored, ii) even trigger size and iii) fixed shape, which are less realistic in practice. In addition, it is unclear whether this method is (computationally) feasible for larger-size triggers with complex patterns, e.g., not only black and white pixels.


NNoculation~\cite{veldanda2020nnoculation} relaxes the assumption of backdoor triggers. To correct the backdoored model, it has two phases: pre-deployment and post-deployment phases. The pre-deployment retrains the (backdoored) DNN model $F_{\Theta_{bd}}$ using an augmented dataset consisting of clean samples and noisy version of clean samples---random noises are added to the clean samples. The latter noisy samples are attempting to approximately approach the poisoned data. In other words, some noise may be able to have, to some extent, trigger effect. Therefore, the pre-deployment constructs an augmented backdoored model $F_{\Theta_{aug}}$, which could reduce the ASR of the attacker. In post-deployment, any input will be fed into both $F_{\Theta_{bd}}$ and $F_{\Theta_{aug}}$, and the input is rejected and then isolated/quarantined on the condition that the predictions between $F_{\Theta_{bd}}$ and $F_{\Theta_{aug}}$ mismatch. In other words, some trigger inputs are misclassified by the $F_{\Theta_{bd}}$ to the attacker chosen class but are still correctly classified by the $F_{\Theta_{aug}}$ to their ground-truth classes. The quarantined data are more likely to include the attacker's trigger instances. Once many quarantined instances are collected, the defenders train a CycleGAN~\footnote{The CycleGAN~\cite{zhu2017unpaired} performs the image-to-image transformation, where images from one distribution are transformed into another image distribution.} that is used to transform between clean inputs and poisoned inputs---the CycleGAN learns to stamp triggers to clean inputs. Using those generated poisoned images and their corrected labels, the defender can retrain the $F_{\Theta_{bd}}$ to reduce the likelihood of the effectiveness of the actual backdoor. There are several limitations to this work. Firstly, like NeuralCleanse, it is mainly applicable for the image domain and could result in high computational cost. Secondly, it requires professional ML expertise. Thirdly, the selection criteria, e.g., the number of quarantined samples required, are mostly empirically based. Inadvertently, the trigger input is more likely to successfully bypass the system at the early deployment stage as this system needs to capture a number of quarantined data. Moreover, it is not always stable~\cite{sarkar2020facehack}.


\subsection{Discussion and Summary}
Table~\ref{tab:countermeasure} compares different countermeasures. It is under the expectation that none of them can defend all backdoor attacks, all with their limitations. For examples, Activation clustering~\cite{chen2018detecting}, spectral signature~\cite{tran2018spectral} and SCAn~\cite{tang2019demon} are offline data inspection method and assume access to the trigger inputs, therefore, it is only applicable to \texttt{data collection} attacks. For NeuralCleanse~\cite{wangneural} and TABOR~\cite{guo2019tabor} that reverse-engineering the trigger, besides the acknowledged high computational cost and only work for small trigger size~\cite{wangneural}, they cannot be applied to binary classification tasks such as malware detection---classes shall be higher than two~\cite{xu2019detecting}. The computational cost could be quite high when the number of classes is large.

Most countermeasures need a (small) set of validation data that contains no trigger, which renders difficult under \texttt{collaborative learning} attacks. It is also noticeable that some countermeasures rely on global reference---in most cases, training shadow models---to determine a threshold to distinguish whether the model is clean. This results in two inadvertent limitations: the global threshold cloud be unreliable to some unique tasks; obtaining this global reference could require high computational resources, e.g., to train (many) shadow models or/and ML expertise. The latter appears cumbersome against \texttt{outsource} and \texttt{pretrained} attacks, where the defender is usually limited by computational resource or/and ML expertise. Therefore, it is essential to consider the defender's resources---\cite{xu2019detecting,kolouri2020universal} otherwise requires training a large number of shadow models by the defender. When devising countermeasures, assumptions should be made reasonable---\cite{xu2019detecting,qiao2019defending} otherwise assumes the trigger size is known. For online inspection, it is worth mentioning again that it does require preliminary offline preparation. It is also imperative to reduce the detection latency, as online inspection is performed during run-time, to be mountable for real-time applications.

All of the backdoor countermeasures are empirically based except two concurrent provable works~\cite{wang2020certifying,weber2020rab} that both utilize randomized smoothing---a certification technique. However, the major limitation of certification is that it always assumes ($l_p$-norm) bounded perturbation of the trigger. Such an assumption is not always held for backdoor triggers. The trigger can be perceptible that is unbounded but still inconspicuous. Moreover, the computational overhead of \cite{weber2020rab} may not be acceptable in some scenarios. We regard that it is hugely challenging to devise practical provable robustness defense against backdoor attacks, though it is preferable.

It is worth mentioning that several countermeasures against backdoor attacks are applicable to defend other adversarial attacks. The backdoor attack is akin to UAP; hence, defending against the backdoor attack would be likely to defend against the UAP. The SentiNet~\cite{chou2018sentinet} has shown that their defending method could be generalized to both backdoor attack and UAP. The NIC~\cite{ma2019nic} also shows that it is effective against UAP.

We also recommend the defense studies explicitly identify threat models and clarify the defender's capability. It is better to specify which attack surfaces are concerned. In addition, it is encouraged that the defense always considers backdoor variants and adaptive attacks, if possible---we do recognize that defending adaptive attack appears to be an open challenge.

\section{Flip Side of Backdoor Attack}\label{sec:flipside}
Every coin has two sides. From the flip side, backdoor enables several positive applications.

\subsection{Watermarking}
Watermarking of DNN models leverages the massive overcapacity of these models and their ability to fit data with arbitrary
labels~\cite{szyller2019dawn}. There are current works considering backdoor as a watermark to protect the intellectual property (IP) of a trained DNN model~\cite{chen2018blackmarks,li2019persistent,adi2018turning,guo2018watermarking,zhang2018protecting,darvish2019deepsigns}. The argument is that the inserted backdoor can be used to claim the ownership of the model provider since only the provider is supposed to know such a backdoor. In contrast, the backdoored DNN model has no (or imperceptible) degraded functional performance on normal inputs. Though there are various countermeasures---detection, recovery, and removal---against backdoor insertion, we conjecture that the watermark using the backdoor techniques could be often robust. The rationale is that it is very challenging to develop one-to-all backdoor countermeasures. However, we do recommend that careful backdoor insertion strategy should always be considered to watermark the model, especially considering the adaptive backdoor insertion. It is feasible to utilize backdoor as an information hiding technique~\cite{kong2019adversarial} or stenography technique.


\subsection{Against Model Extraction}
Jia \textit{et al.}~\cite{jia2020entangled} proposed Entangled Watermarking Embeddings (EWE) to defend against model extraction attacks~\cite{tramer2016stealing}. The EWE is an extension of the watermarking but for different attack models. Here, the attacker aims to steal the models provided by a vendor, e.g., machine learning as a service (mlaas), using model extraction attacks, generally relying on querying the victim model and observing the returned response, e.g., softmax. This attacking process is akin to chosen-plaintext attacks in cryptography. Then the attacker does not need to pay the victim model provider for further queries because the stolen model has comparable accuracy of the victim model. The attacker may also seek to make a profit by publishing the stolen model to the public. 

The EWE identifies the overfitting weakness of the traditional watermarking method. This means the model parameters in charge of the main task and watermark (backdoor) task are separated. Therefore, when the attacker queries the model aiming to steal the main task's functionality, the watermark as a different sub-task may not be propagated to the stolen copy. Accordingly, the EWE utilizes a method to jointly learn how to classify samples from the main task distribution and watermarks (sub-task) on vision and audio datasets. In other words, the backdoor inserted by the model provider will be unavoidably propagated to the stolen model. The advantage of EWE lies in its minimal impact on the model's utility, particularly less than 1\% CDA drop. In addition, it is a challenge to remove the backdoor by the attacker unless the attacker is willing to sacrifice the CDA performance on legitimate data inevitably. Note the later violates the purpose of model extraction that is to steal a well-performing model. Moreover, the EWE is designed in a way to claim the ownership of the model provider with high confidence, e.g., 95\%, in just {\it a few} (remote) queries, e.g., 10.

\subsection{Against Adversarial Examples}

Shan {\it et al.}~\cite{shan2019using} exploit the backdoor as an intentionally designed `trapdoor' or `honeypot' to trap an attacker to craft adversarial examples that will fall in the trapdoor, then be captured. This method is fundamentally different from previous countermeasures against adversarial examples that neither tries to patch nor disguise vulnerable points in the manifold. Overall, trapdoors or backdoors are proactively inserted into the models. At the same same, the neuron activation signatures corresponding to trapdoors are extracted. When the attacker attempts to craft adversarial examples, usually relying on optimization, these optimization algorithms will toward trapdoors, leading them to produce attacks similar to trapdoors in the feature space. Therefore, the adversarial examples can be detected with high detection success rate and negligible impact on normal inputs---does not impact normal classification performance. The advantage of the defender now is that adversarial attacks are more predictable because they converge to a known region (known weakness of the model that is the trapdoor) and are thus easier to detect. This method is easy to implement by simply implant backdoor into the model before deployment; the attacker will have little choice but to produce the ``trapped'' adversarial examples even in the white-box setting. This strategy could be applicable to trap universal adversarial patch.

\subsection{Data Deletion Verification}
Sommer {\it et al.}~\cite{sommer2020towards} propose to utilize backdoor effect to verify the user data deletion when the data deletion is requested by the user. For example, different users contribute their data to the server. The server trains a model over the contributed data. However, in some cases, the user may ask for the revocation of the data contribution. In this case, the server needs to unlearn the data from this user. One intuitive way is to retrain the model from scratch after removing the user data as requested. For the requested user, it is non-trivial to verify this deletion performed by the server. In this context, Sommer {\it et al.}~\cite{sommer2020towards} constructively turn the backdoor trace as a means of verifying such an data deletion operation. Generally, when the user contributes the data, a fraction of the data is poisoned with the user secret chosen trigger. When the server model is trained on the user data, there is a backdoor trace left. If the data deletion is performed, then such backdoor trace should not exist. The backdoor trace can be checked by the ASR that should be very low given the data deletion is complied. Because the model now should be trained without the user data, neither the backdoor trace contained in the data. We note that Sommer {\it et al.} use the conventional data poisoning technique in order to insert backdoor trace. In other words, the label is changed to the target label given the poisoned data samples. This leads to the inconsistency between the label and data content, which could be recovered even when the data undergoes manual/visual inspection. It is preferable to adopt data poisoning techniques devised under \texttt{data collection} attack (\autoref{sec:TM3Attack}) to keep the label and content being consistent. Though \texttt{data collection} attack usually requires knowledge of the model architecture used later on, it is reasonable to gain such knowledge under the data deletion verification application, where the server shall inform such knowledge to data contributor.




\section{Discussion and Prospect}\label{sec:discussion}
\subsection{Adaptive Attack}
We conjecture that, akin to adversarial
examples, there will be a continuous arms race for backdoor attacks. New devised empirical countermeasures against backdoor attacks would soon be broken by strong adaptive backdoor attacks when the attacker is aware of the defense. In this case, it is not hard to bypass the defense, which has been shown by taking the evasion of defenses such as spectral signature~\cite{tran2018spectral}, activation clustering~\cite{chen2018detecting}, STRIP~\cite{gao2019strip}, and NeuralCleanse~\cite{wangneural} as objectives into the loss function when inserting the backdoor to the model~\cite{bagdasaryan2018backdoor,costales2020live,tan2019bypassing,bagdasaryan2020blind}. Though defenses can increase the cost paid by the attackers, fundamentally preventing backdoor attacks is regarded as a daunting problem.

\subsection{Artifact Release}

Recently, there is a trend of encouraging artifacts release, e.g., source code or tools, to facilitate the reproduction of each study. Fortunately, many studies follow such a good practice, and the link is attached in each corresponding reference given that the artifacts are released---most of them are released at \texttt{Github}. It is encouraged that the following works in this research line always well document and publish their artifacts. There are efforts to encapsulate different works into a toolbox~\cite{karra2020trojai,nicolae2018adversarial} to ease the rapid adoption and facilitate understanding. It is worth highlighting that there is an organized online backdoor detection competition~\url{https://pages.nist.gov/trojai/}.


\subsection{Triggers}
According to the summary in Table~\ref{tab:countermeasure}, it is observed that most countermeasures are ineffective in detecting large size triggers or/and class-specific triggers, which demand more defense efforts. To facilitate backdoor insertion such as data poisoning under \texttt{data collection}~\cite{xi2020graph} as well as attack phase. The trigger can be crafted adaptively as long as their latent representation is similar. In other words, the triggers do not have to be visually universal (same). For instance, imperceptible triggers are used to poison the data while the perceptible trigger is used during an attack to increase the attack robustness~\cite{saha2019hidden}; graph triggers are adaptively tailored according to the input graph, thereby optimizing both attack effectiveness and evasiveness on backdoored graph neural networks~\cite{xi2020graph}. Dynamic triggers are effective regardless of location, visualization variations~\cite{salem2020dynamic,li2020rethinking}.

One most advantage of backdoor is the arbitrarily chosen trigger, e.g., in comparison with adversarial example attacks, which is fully under the control of the attacker. Therefore, the triggers can be optimally identified or chosen to facilitate the attack, especially survive various variations in the physical attack~\cite{pasquini2020trembling}. In fact, various natural triggers have been exploited including natural accessory triggers~\cite{wenger2020backdoor}, facial expression~\cite{sarkar2020facehack}, natural reflection phenomena~\cite{liu2020reflection}.

\subsection{Passive Attack}
We believe that the backdoor attacker needs to modify both the training data and testing data by inserting the trigger in the input. This means the trigger has to be actively added to the input during the attacking phase. In other words, the attacker performs the backdoor attack actively. However, the attacker could select some trigger to carry out a passive attack, referred to as a different term of semantic backdoors~\cite{bonawitz2019towards}. For example, a red hat as a trigger for objection detector. Then the victim could be some pedestrians who accidentally wears a red hat. The backdoored self-driving car will ignore their existence and causes casualties. These victims are innocent personal targeted by the attacker. The other case is when the trigger is some words; whenever the trigger is shown in a review, the review will be classified as negative regardless of the real comments. In the attack phase, there is no need to modify the trigger input by the attacker actively. 

\subsection{Defense Generalization}
As a matter of fact, most countermeasures are explicitly or inexplicitly designed or/and have been mainly focusing on or tailored for vision domain in terms of classification task, see summary in Table~\ref{tab:countermeasure}. There are lack of generic countermeasures that are demonstrated to be effective across different domains such as vision, audio and text. Since all of these domains have been infected by backdoor attacks. In addition, backdoor attacks on other than classification tasks such as sequential model or reinforcement learning~\cite{yang2019design,wangstop,wang2020watch,kiourtitrojdrl}, deep neural graph model~\cite{xi2020graph}, deep generative model~\cite{ding2019trojan}, source processing~\cite{ramakrishnan2020backdoors,schuster2020humpty} also solicit defenses for which current countermeasures focusing on (image) classification task may not be directly applicable~\cite{kiourtitrojdrl}. It is recommended that the defense studies clearly clarify the generalization when devising a countermeasure.

\subsection{Defender Capability}
It is essential to develop suitable defenses for users under given threat models.
In this context, the defense development should consider the users' capability. For example, under \texttt{outsourcing}, the users usually have neither ML expertise nor rich computational resources. It is wise to avoid reliance on reference models, especially those requiring substantial computational resources and ML expertise. Even in transfer learning that is affected by \texttt{pretrained} attack surface, the users are always limited with massive GPU clusters when they expedite build accurate models customized to their scenario. Thus, heavy computational defenses could be unsuitable. In addition, some defenses are with strong assumptions such as the knowledge of trigger size~\cite{kolouri2020universal,qiao2019defending} or require extensively auxiliary model training or/and ML expertise~\cite{xu2019detecting,kolouri2020universal,weber2020rab}, which appear to be impractical or/and too computationally expensive.

However, we do recognize that defenders are with harder tasks and appreciate more studies on this side with reasonable assumptions and acceptable efficacy, in most cases, properly tailored to counter attackers under a given threat model---one to all countermeasure could be an open challenge.

\subsection{In-House Development}
As explicitly suggested~\cite{taddeo2019trusting}, it is a necessity of developing critical security applications by reliable suppliers. Model training and testing as well as data collection and validation shall be performed by the system providers directly and then maintained securely. As such, all threats to backdoor attacks can be extremely avoided, if not all can be ruled out, because the backdoor attack does usually require to tamper the training data or/and model at any context by the attacker. As a comparison, other forms of adversarial attacks in particular adversarial example only needs to tamper the fed input during deployment that is usually out of the control of the system owner/provider, where this in-house application development may become cumbersome. Therefore, the practice of enforcing in-house application will be very useful to reduce attack surfaces for backdoor attacks, especially on critical infrastructures, if it cannot eschew all.

\subsection{Majority Vote}
Unlike the adversarial examples that are transferable among models, backdoor attacks have no such advantage. This is because the model needs to be firstly tampered to insert a backdoor. Therefore, it is worth considering the majority vote decision where several models are used to make decisions collaboratively. The decision will only be accepted when they are the same or meeting with the majority decision criterion. To achieve this goal, we can build models with different sources. To be precise, for outsourcing, the user can outsource the model training to non-colluding parties, e.g., Microsoft and Google. Even though both return backdoored models, it is doubtful that they will choose the same trigger. This is also the case for the pretrained models. The users can randomly choose pretrained models from different sources to perform transfer learning. Similarly, it is doubtful that the attacker will insert the same trigger to all of them. In comparison with in-house development, we regard the majority decision as a more practical strategy against backdoor attacks.

\section{Conclusion}
This work provides a systematic and comprehensive review of the research work about backdoor attacks on deep learning and sheds light on some critical issues in the design of countermeasures to mitigate backdoor attacks. We found that there is still a significant gap between the existing attacks and countermeasures. To develop practical working solutions, we should consider avoiding less realistic assumptions as analyzed and highlighted in this review.


\end{document}